\documentclass[11pt]{article}

\usepackage{amsmath}
\usepackage{graphicx}
\usepackage{indentfirst}
\usepackage{amssymb}
\usepackage{cite}
\usepackage{color}
\usepackage{subfigure}
\usepackage{varwidth}
\usepackage{subfigure}

\usepackage[colorlinks=true, linkcolor=red, citecolor=blue, urlcolor=magenta]{hyperref}

\begin{document}
\title{ Optical appearance of the Konoplya-Zhidenko rotating non-Kerr black hole surrounded by a thin accretion disk }

\date{}
\maketitle

\begin{center}
\author{Ke-Jian He,}$^{a}$\footnote{E-mail: kjhe94@163.com}
\author{Yi Liu,}$^{a}$\footnote{E-mail: yiliuphysics@163.com}
\author{Chen-Yu Yang,}$^{a}$\footnote{E-mail:  chenyu\_yang2024@163.com}
\author{Xiao-Xiong Zeng}$^{b}$\footnote{E-mail: xxzengphysics@163.com (Corresponding author)}
\\

\vskip 0.25in
$^{a}$\it{Department of Mechanics, Chongqing Jiaotong University, Chongqing 400000, People's Republic of China}\\
$^{b}$\it{College of Physics and Electronic Engineering, Chongqing Normal University, Chongqing 401331, People's Republic of China}\\
\end{center}
\vskip 0.6in
{\abstract
{ In this study, we investigate the optical appearance of rotating Konoplya-Zhidenko non-Kerr black holes in the presence of thin accretion disks, with the aim of examining whether the information of deformation parameters manifest in observable signatures. By employing a fisheye camera model in conjunction with backward ray-tracing techniques, we simulate images for both prograde and retrograde accretion scenarios. The results indicate that the deformation parameter $\xi$ can partially mitigate the shadow deformation induced by the rotation parameter $a$. The inner shadow displays characteristic morphological transformations at varying observation angles, transitioning from an axisymmetric circular form at low angles to a hat-like configuration at higher angles. Furthermore, at high observational inclination angles, the direct image and the lensed image become distinctly discernible, and an increase in the deformation parameter $\xi$ enhances the observed intensity of the image. Interestingly, the motion behavior of the accretion flow influences the observed intensity distribution on the screen, a finding that is consistent with the redshift distribution. Hence, variations in the deformation parameter $\xi$, the observation angles, and the motion behavior of the accretion flow collectively influence the observable appearance of the black hole. We expect this work to provide valuable references for identifying observable signatures of spacetime deviations from general relativity.}}

\thispagestyle{empty}
\newpage
\setcounter{page}{1}

\section{Introduction}
\label{sec:intro}
In accordance with the theory of general relativity (GR), extremely dense celestial bodies, including black holes, are known to exist within our universe, and the quest to identify these entities has become a focal point of research in astronomy and astrophysics. In 2016, the LIGO-Virgo collaboration's detection of the gravitational wave event GW150914, produced by the merger of two black holes with masses of 29 $M_{\odot}$ and 36 $M_{\odot}$ (where $M_{\odot}$ denotes the solar mass), provided robust empirical evidence for the existence of these enigmatic astrophysical objects \cite{LIGOScientific:2016sjg,LIGOScientific:2016vlm,LIGOScientific:2016aoc}.
Very recently, the Event Horizon Telescope (EHT) has played a pivotal role in capturing images of the supermassive black hole at the center of the M87*, utilizing 1.3 mm interferometric observations\cite{EventHorizonTelescope:2019dse,EventHorizonTelescope:2019uob,EventHorizonTelescope:2019jan,EventHorizonTelescope:2019ths,
EventHorizonTelescope:2019pgp,EventHorizonTelescope:2019ggy}. Subsequently, the EHT unveiled an image of Sagittarius A* (Sgr A*), which not only confirmed the presence of a supermassive black hole at the center of the Milky Way galaxy but also validated the analysis of stellar trajectories \cite{EventHorizonTelescope:2022wkp,EventHorizonTelescope:2022vjs,EventHorizonTelescope:2022wok,EventHorizonTelescope:2022exc,EventHorizonTelescope:2022urf,EventHorizonTelescope:2022xqj} .
In the images of the EHT, it is evident that there is a bright circular structure, which is interpreted as the photon ring, and a central dark region, referred to as the black hole shadow.
Initial approaches to observing the black hole photonsphere and shadow were presented in\cite{Virbhadra:1999nm,Claudel:2000yi,Virbhadra:2007kw,Falcke:1999pj}; since then, the methodologies for computing these shadows have undergone continuous improvement \cite{Perlick:2021aok}. At present, the  black hole shadow has been widely studied in many fields, including but not limited to serving as a standard scale to quantify and test various cosmological phenomena \cite{Tsupko:2019pzg,Vagnozzi:2022moj,Chen:2022nbb}, quantify and constrain black hole parameters  \cite{Hioki:2009na,Li:2013jra,Wei:2018xks,Younsi:2016azx,Konoplya:2021slg,Tsukamoto:2014tja,yang2026distinguishing2,Zhong:2021mty,Huang:2024wpj,Wang:2024uda}, explore fundamental physical issues such as dark matter and quasinormal modes\cite{Konoplya:2019xmn,Hou:2018bar,Konoplya:2019sns,Haroon:2018ryd}, and test various gravitational theories \cite{yang2025observational2,Mou:2022tln,He:2024mal,He:2024bll,yang2026observational,He:2024yeg,Abdujabbarov:2016hnw,Zeng:2021dlj,Atamurotov:2015nra,Perlick:2015vta,Atamurotov:2021cgh,Wang:2022kvg,Chen:2022scf,
He:2024qka,Guo:2020zmf,Atamurotov:2013sca,Shaikh:2018lcc,yang2026shadow,Qu:2023hsy,Papnoi:2014aaa,Meng:2023wgi,Zhang:2023bzv,Zhu:2021tgb,He:2025qmq,yang2026distinguishing}.

On the other hand, the luminous structures observed in the image are generated by the light emanating from materials within the accretion disk surrounding the black hole. In 1979, Luminet considered  a model of a thin accretion disk, utilized semi-analytical techniques to simulate the image of a Schwarzschild black hole, and demonstrated that the appearance of the black hole's shadow is contingent upon the characteristics of the accretion flow and  the outer edge of central brightness depression\cite{Luminet:1979nyg}. Considering a relatively simple accretion model, spherical accretion model, some meaningful results are also obtained \cite{Narayan:2019imo,Zeng:2020dco,Heydari-Fard:2023ent}.
In the work of\cite{Gralla:2019xty}, Gralla et al. examined Schwarzschild black holes surrounded by a geometrically and optically thin accretion disk, revealing that the bright ring outside the black hole's shadow consists of direct emissions, a lensing ring, and a photon ring. Subsequently, the model proposed by Gralla et al. was extended to encompass additional scenarios involving matter fields and modified gravity; however, these investigations were confined to the context of spherically symmetric black holes\cite{Zeng:2020vsj,Peng:2020wun,He:2022yse,Li:2021ypw,Zeng:2021mok,Li:2021riw,Gao:2023mjb,Cui:2024wvz,He:2021htq,Wang:2022yvi}. Given the strong magnetic field surrounding the black hole, the thin accretion disk model is extended to the case of a rotating Kerr-Melvin black hole, which is composed of plasma fluids and behaves differently in different regions\cite{Hou:2022eev}. This research investigates the utilization of internal shadows and critical curves of a black hole to estimate the magnetic field surrounding it. By considering the thin disk model presented in the work\cite{Hou:2022eev}, several intriguing studies have also been conducted on alternative models of black holes within modified gravitational backgrounds\cite{Yang:2024nin,He:2024amh,Li:2024ctu}. By building on these previous studies, the choice of a geometrically thin accretion disk model in the present work is motivated by both its theoretical advantages and its astrophysical relevance. Its simple emission geometry and relatively mature theoretical framework make it possible to characterize more directly the effects of the background spacetime on particle motion, photon propagation, and shadow formation, while minimizing the influence of disk thickness, complicated magnetohydrodynamic processes, and detailed radiative transfer effects \cite{ShakuraSunyaev1973,NovikovThorne1973,PageThorne1974}. From an astrophysical perspective, geometrically thin disks are also well motivated in black hole X-ray binaries during the thermal state, where the observed continuum emission is commonly modeled by the Novikov-Thorne thin disk framework and used, for example, in black hole spin measurements \cite{RemillardMcClintock2006,Kulkarni2011,McClintock2014}. Therefore, although the present model is idealized, it still provides a clean and effective benchmark for isolating the observable imprints of the background spacetime and facilitates comparison with previous theoretical studies.

It is well known that the current observational results cannot completely rule out the possibility of deviating from GR, thus leaving an open window for testing other modified gravity theories. By incorporating static deformations, Konoplya and Zhidenko constructed a rotating non-Kerr black hole solution beyond the framework of GR \cite{Konoplya:2016pmh}, which can be interpreted as an axisymmetric vacuum solution of an unspecified alternative theory of gravity \cite{Konoplya:2016jvv}. In contrast to the Kerr metric, which is characterized solely by mass and rotation parameter, the Konoplya-Zhidenko rotating non-Kerr black hole incorporates an additional deformation parameter. This deformation parameter characterizes the deviation from the standard Kerr black hole solution and substantially modifies the spacetime geometry in the strong-field regime \cite{Wang:2016paq}. By employing this rotating non-Kerr metric, it is demonstrated that some non-negligible deformations of the Kerr spacetime can lead to the same frequencies of the black-hole ringing\cite{Konoplya:2016pmh}, as evidenced by data from GW150914\cite{LIGOScientific:2016aoc,LIGOScientific:2016lio}. Furthermore, the constraints imposed by quasi-periodic oscillations and iron line profiles\cite{Bambi:2012pa,Bambi:2013fea,Kong:2014wha,Bambi:2011ek,Johannsen:2012ng,Jiang:2014loa,Bambi:2013sha} lend additional support to the hypothesis that actual astrophysical black holes can be accurately characterized using the Konoplya-Zhidenko rotating non-Kerr metric. Several intriguing studies have explored the properties and characteristics of the Konoplya-Zhidenko rotating non-Kerr black hole spacetime, encompassing aspects such as energy extraction, strong gravitational lensing effects, and magnetic reconnection, among others\cite{Long:2017xqr,Cunha:2017eoe,Long:2024tws,He:2023ife,Patra:2024srh}.

Due to the asymptotic behavior of the Konoplya-Zhidenko rotating non-Kerr black hole closely resembling that of the Kerr black hole\cite{Wang:2016paq}, it is challenging to differentiate between the two in weak-field regimes, such as those observed within the solar system. Therefore, observations in strong-field regimes can yield valuable insights into potential deviations from the Kerr solution, with the black hole shadow providing an ideal observational platform for such tests. Indeed, within the Konoplya-Zhidenko rotating non-Kerr black hole spacetime background, novel properties of the spacetime structure induced by the deformation parameter have been identified, leading to significant implications for the morphology of the black hole shadow.
In \cite{Wang:2017hjl}, Wang et al. investigated the shadow cast by the Konoplya-Zhidenko rotating non-Kerr black hole, the results  show that the black hole shadow exhibited a spiky feature with minor eyelash-like structures in the case of $a>M$. As the deformation parameter increased, the spiky characteristic of the shadow progressively diminished. Subsequently, the shadow properties of the Konoplya-Zhidenko naked singularity were further investigated \cite{Wang:2023jop}, revealing that these naked singularities can produce both arc-like and ring-shaped shadows under rotational conditions. However, both of these studies on the shadow of Konoplya-Zhidenko rotating non-Kerr black hole were limited to the assumption of a celestial sphere light source and did not incorporate the accretion disk model.
Hence, our aim is to further investigate the optical observational characteristics of Konoplya-Zhidenko rotating non-Kerr black holes surrounded by thin accretion disks. In this scenario, it is possible to obtain more comprehensive physical information, including the Doppler effect, redshift, and the kinematics of accretion flows.  In this study, we systematically investigated the influence of the motion behavior of accreted matter and the variation of the observed inclination angle on the optical observational characteristics of black holes. Additionally, we investigate whether the characteristics of spacetime deformation parameters are discernible in the optical appearance of black holes.

The paper will be completed in the following manner. In Section 2, we will provide a concise introduction to the Konoplya-Zhidenko rotating non-Kerr black hole.  Moreover, the geodesic motion behavior of photons in this spacetime was discussed. In Section 3, we study the shadow of the black hole, and analyzed the deviation from the circularity and the size of shadow. Also, the shadow angular diameter is obtained, and compared with that of M87* and SgrA*. In Section 4, we examine an optically and geometrically thin accretion disk situated on the equatorial plane of the black hole to derive its visually observed appearance. We primarily focus on the influence of variations in the deformation parameter, the observed inclination angle and the accretion flow behavior on the resulting observational characteristics. In Section 5, we meticulously scrutinize the impacts of parameters on the redshift factor and the appearance of the inner shadow. Finally, in Section 6, we give a brief conclusion and discussion.

\section{The  Konoplya-Zhidenko rotating non-Kerr black hole and null geodesic}
\label{sec2}
As a starting point, we will present a succinct overview of the Konoplya-Zhidenko rotating non-Kerr spacetime, which serves as the foundational framework for our comprehensive investigation. The metric presented here, which deviates from the Kerr metric, is a non-Kerr metric that typically arises from deforming the Kerr metric. In the Boyer-Lindquist coordinates, the metric of a rotating non-Kerr black hole in the Konoplya-Zhidenko model can be expressed as\cite{Konoplya:2016pmh}
\begin{align}\label{metric1}
&ds^2=-\frac{\mathcal{A}^2(r,\theta)-\mathcal{B}^2(r,\theta)\sin^2\theta}{\mathcal{C}^2(r,\theta)}dt^2-2\mathcal{B}(r,\theta)r\sin^2\theta dt d\varphi+\mathcal{C}^2(r,\theta)r^2 \sin^2\theta  d\varphi^2 \nonumber\\
&~~~~~~~~+\Sigma(r,\theta)r^2 d\theta^2+\frac{\Sigma(r,\theta)\mathcal{D}^2(r,\theta)}{\mathcal{A}^{2}(r,\theta)}dr^2
\end{align}
in which
\begin{align}\label{metric2}
&\mathcal{A}^2(r,\theta)=\frac{r^2+a^2-2Mr}{r^2}-\frac{\xi}{r^3},  \qquad \mathcal{B}(r,\theta)=\frac{2Ma}{r^2+a^2\cos^2\theta}+\frac{a\xi}{r^2(r^2+a^2\cos^2\theta)}, \nonumber\\
&\mathcal{C}^2(r,\theta)=\frac{(r^2+a^2)^2-a^2\sin^2\theta(a^2+r^2-2Mr)}{r^2(r^2+a^2\cos^2\theta)}+\frac{a^2 \xi \sin^2\theta}{r^3(r^2+a^2\cos^2\theta)}, \nonumber\\
&\mathcal{D}(r,\theta)=1, \qquad  \Sigma(r,\theta)=\frac{a^2\cos^2\theta+r^2}{r^2}.
\end{align}
In this spacetime, the mass of the black hole is denoted by $M$, while the rotation parameter is represented by $a$. Additionally,  the deviation from Kerr spacetime is quantified by the deformation parameter $\xi$, which degenerates to the general Kerr black hole metric at $\xi=0$. It is worth mentioning that the metric \ref{metric1} has an asymptotic behavior similar to that of the Kerr metric at infinity, but the existence of the deformation parameter $\xi$ alters the nature of the spacetime at the near-horizon region. In other words, the deformation parameter plays a crucial role in the dynamical behavior of the event horizon, as can be seen from the definition equation of the event horizon of a black hole, that is,
\begin{align}\label{EH}
\Delta_r=r^2-2Mr+a^2-\frac{\xi}{r}=0.
\end{align}
Obviously, the location of the event horizon $r_h$ is evidently distinct from that of the Kerr black hole, and it explicitly relies on the value of the deformation parameter $\xi$. For the value of the deformation parameter $\xi$, it can be classified as
\begin{align}
    \begin{cases}\xi \geq \xi_{c1} \equiv -\frac{2}{27}(\sqrt{4M^2-3a^2}+2M)^2(\sqrt{4M^2-3a^2}-M), &  a<M \\
     \xi>0, & a>M \label{EQ4.5}
    \end{cases}
\end{align}
which allows for the existence of an event horizon in this specific black hole spacetime, and further details can be found in Refs.\cite{Wang:2017hjl,Wang:2023jop}.  The absence of a horizon occurs when $\xi$ and $a$ lie in other regions, resulting in the emergence of a naked singularity within the spacetime. In light of the Weak Cosmic Censorship Conjecture (WCCC), our exclusive focus will be on scenarios where the existence of the event horizon is ensured, and the values of the relevant parameters adhere to this principle. Nevertheless, outside the parameter region admitting a positive real root of $\Delta_r=0$, the Konoplya-Zhidenko spacetime describes a naked singularity. Previous work has shown that Konoplya-Zhidenko naked singularities can still possess complete or incomplete photon sphere structures in certain parameter ranges \cite{Wang:2023jop}. More generally, studies of other naked singularity spacetimes suggest that strongly naked singularities without photon spheres can exhibit qualitatively different optical appearances \cite{gyulchev2020observational}, and that thin accretion disk images may help distinguish black holes from naked singularities \cite{shaikh2019can}. These cases are not considered in the present work and will be addressed in future work.

The Hamiltonian of particle propagation along geodesics in this spacetime is given by
\begin{align}\label{HJ}
\mathcal{H}=\frac{1}{2} g^{\mu\nu} p_\mu p_\nu=-\frac{u^2}{2}.
\end{align}
Where $\mathcal{H}$ represents the Hamiltonian, $p_\mu$ denotes the four-momentum of the particle, and $u$ signifies the mass of  particles moving in the black hole spacetime, with $u=0$ indicating null geodesics. By combining the spacetime metric (\ref{metric1}) with the Hamiltonian (\ref{HJ}), one can find that the Hamiltonian $\mathcal{H}$ being independent of $t$ and $\varphi$ gives rise to two Killing vector fields, $\partial_t$ and $\partial_\varphi$, which represent the translational and rotational invariance of time. Consequently, this leads to the existence of two conserved quantities for photon motion in the Konoplya-Zhidenko rotating black hole spacetime, which are
\begin{align}\label{CQ}
E=-p_t=-g_{\varphi t}\dot{\varphi}-g_{t t}\dot{t}, \qquad  L=p_\varphi=-g_{\varphi \varphi}\dot{\varphi}+g_{\varphi t}\dot{t}.
\end{align}
In the above equations, the dot denotes derivative with respect to the affine parameter $\lambda$, while $E$ and $L$ represent the energy and angular momentum along the axis of symmetry. With the aid of these two conserved quantities, the null geodesics of a photon in this particular spacetime can be formulated as a set of four first-order differential equations, that is
\begin{align}\label{GE1}
\dot{t}=E+\frac{(a^2 E-a L +r^2 E)(2M r^2 +\xi)}{r^3 \Sigma(r,\theta) \mathcal{A}^2(r,\theta)},
\end{align}
\begin{align}\label{GE2}
\dot{\varphi}=\frac{a E \sin^2\theta (2Mr^2+\xi)+a L r \cos^2\theta-L(2Mr^2-r^3+\xi)}{r^3 \Sigma(r,\theta) \mathcal{A}^2(r,\theta)\sin^2\theta},
\end{align}
\begin{align}\label{GE3}
\Sigma^2(r,\theta) r^2 \dot{r}^2=\mathbf{R}(r)=-\mathcal{A}^2(r,\theta)(Q_\kappa+(aE-L)^2)+{(aL-(a^2+r^2)E)^2},
\end{align}
\begin{align}\label{GE4}
\Sigma^2(r,\theta) r^4 \dot{\theta}^2=\Theta(\theta)=Q_\kappa+a^2E^2\cos^2\theta-\frac{\cos^2\theta L^2}{\sin^2\theta}.
\end{align}
where the quantity $Q_\kappa$ represents the third conserved constant of photon motion in this particular spacetime, analogous to the Carter constant associated with the Kerr spacetime\cite{Wang:2017hjl}.
 It is well known that the shadow boundary of a black hole is uniquely determined by its photon region, which always satisfies the conditions $\dot{r}=0$ and $\ddot{r}=0$,  indicating that
\begin{align} \label{PS1}
\mathbf{R}(r) =0, \quad \mathbf{R}'(r)=0
\end{align}
In addition, the unstable photon orbit also needs to meet the condition
\begin{align} \label{PS3}
\mathbf{R}''(r)>0.
\end{align}
The above equations accurately characterize the motion behavior of photons in this spacetime and serve as the foundation for investigating the shadow of the  Konoplya-Zhidenko rotating non-Kerr black hole.

\section{The deformation of the shadow and the angular diameter}
\label{sec3}
\begin{figure}[htbp]
	\centering
	\includegraphics[width=0.35\linewidth]{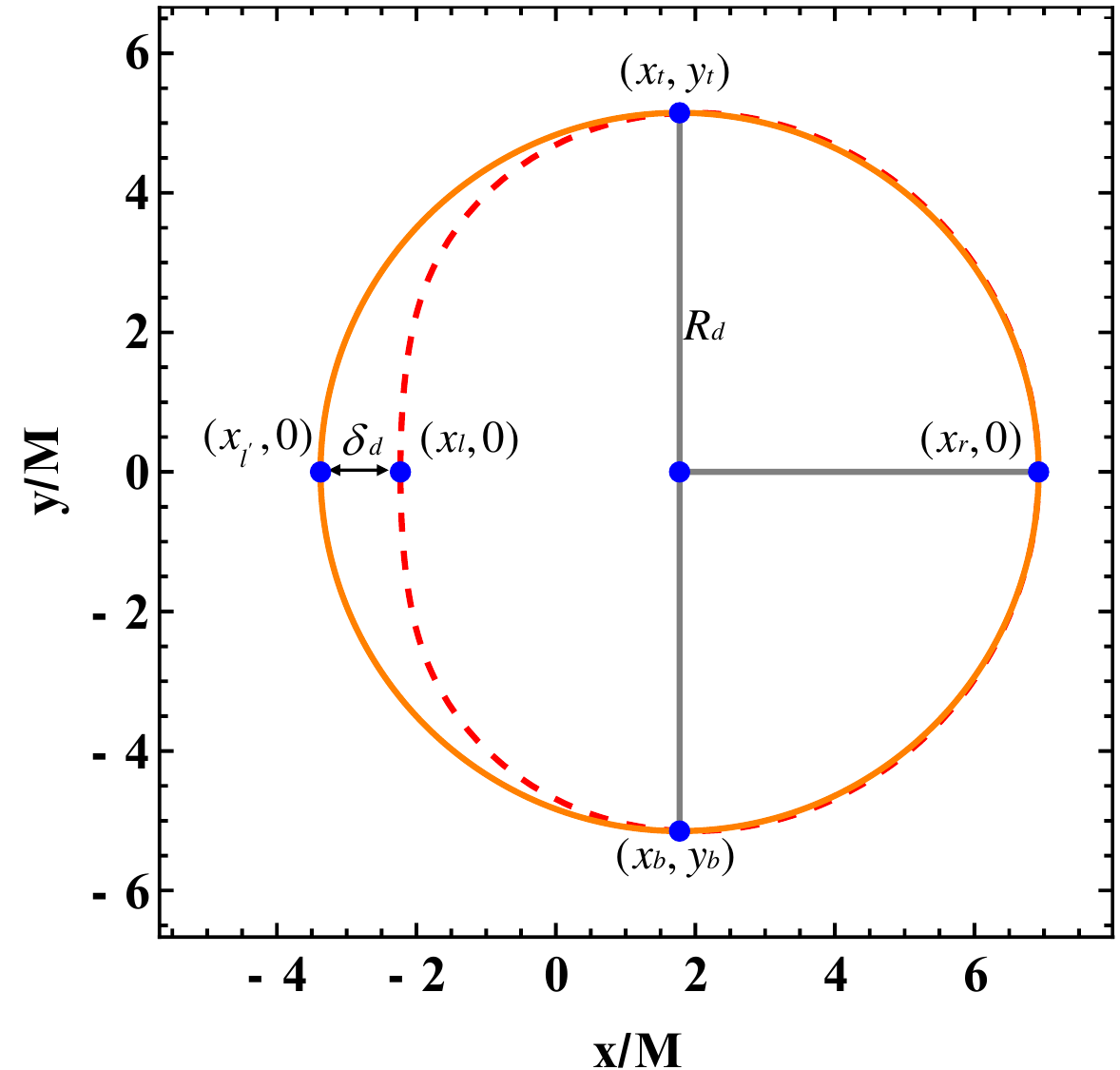}
	\caption{The black hole shadow and the reference circle. In the figure, the red outline represents the black hole shadow, and the orange outline represents the reference circle. The radius of the reference circle is $R_d$, and the absolute horizontal difference between the leftmost points of the reference circle and the black hole shadow is $\delta_d$.}\label{rc}
\end{figure}

To accurately characterize the impact of variations in relevant parameters on the size and deformation of the black hole shadow, two observable physical quantities can be defined, as introduced in \cite{Hioki:2009na}, namely the deviation from the circularity $\delta_d$ and the size $R_d$ of the shadow. The specific form of expression is
\begin{equation}
	R_d = \frac{(x_t - x_r)^2 + y_t^2}{2|x_t - x_r|},\quad \delta_d = \frac{|x_{l^{\prime}} - x_l|}{R_d}. \label{dcs}
\end{equation}
The top, bottom, and rightmost points of the black hole shadow uniquely determine a reference circle, whose radius is denoted as $R_d$, approximating the size of the shadow. The parameter $\delta_d$ represents the absolute horizontal difference between the leftmost points of the black hole shadow and the reference circle, indicating the degree of deviation from circularity, see Figure \ref{rc}.

\begin{figure*}[htbp]
	\centering
	\subfigure[$a=0.3$]{\includegraphics[width=0.35\textwidth]{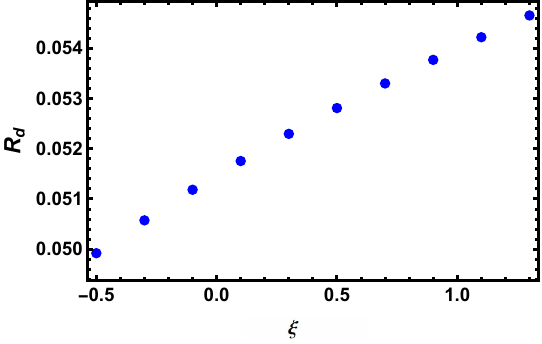}}
	\subfigure[$a=0.3$]{\includegraphics[width=0.35\textwidth]{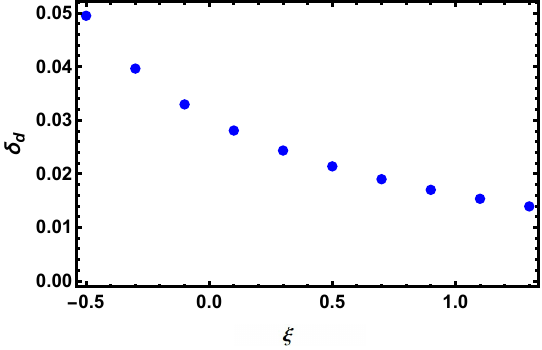}}
	\subfigure[$a=1.15$]{\includegraphics[width=0.35\textwidth]{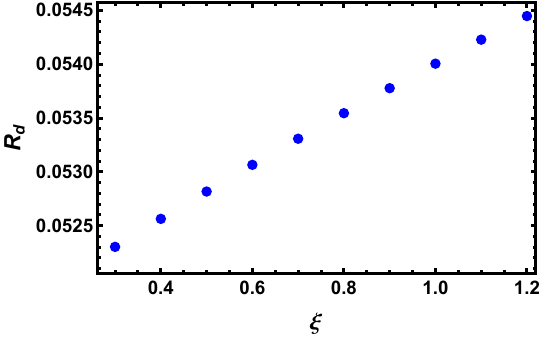}}
	\subfigure[$a=1.15$]{\includegraphics[width=0.35\textwidth]{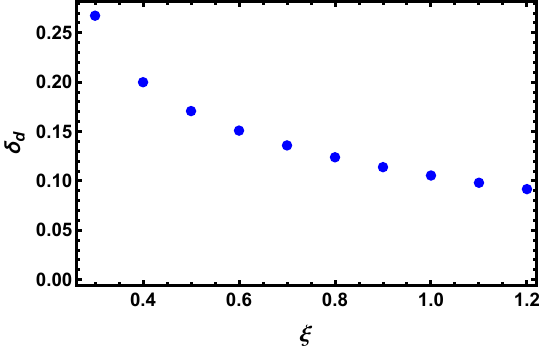}}
	\caption{\label{Rs} The observable $R_d$ and $\delta_d$ for $r_o=100$, $\theta_o=\pi/2$ and $M=1$.}
\end{figure*}

Specifically, the five reference points $(x_t, y_t)$, $(x_b, y_b)$, $(x_r, 0)$, $(x_l, 0)$, and $(x_{l^{\prime}}, 0)$ correspond to the  top, bottom, rightmost, and leftmost points of the shadow, leftmost point of the reference circle,  respectively. When $x_l \neq x_{l^{\prime}}$, $\delta_d \neq 0$. A larger $\delta_d$ indicates a greater deviation of the black hole shadow boundary from a circular shape. In Figure~\ref{Rs}, we present the effects of the deformation parameter $\xi$ and the rotation parameter $a$ on $R_d$ and $\delta_d$. The results demonstrate that for both $a<M$ and $a>M$, as $\xi$ increases, $R_d$ increases while $\delta_d$ decreases, suggesting that the shadow becomes larger and closer to a circular shape.  The observed trends in $R_d$ and $\delta_d$ can be interpreted through the influence of the deformation parameter $\xi$ on the spacetime geometry. It is crucial to note that in the limit of  $\xi=0$, the Konoplya-Zhidenko rotating non-Kerr metric reduces to the standard Kerr metric, providing a clear baseline for comparison.   An increase in  $\xi$ effectively enhances the gravitational strength in the near-horizon region, thereby expanding the photon capture cross-section and resulting in a larger shadow size ($R_d$ increases) compared to its Kerr counterpart with the same spin parameter $a$.  Concurrently, the positive deformation $\xi$ counteracts the frame-dragging effect induced by the black hole's rotation. This restoration of symmetry in the gravitational field reduces the distortion of the contour of the shadow, causing the shadow to appear more circular  ($\delta_d$ decreases) and thus mitigating the oblateness typically induced by rotation in a Kerr black hole.  Conversely, an increase in the rotation parameter $a$ amplifies the frame-dragging effect, which compresses the shadow and enhances its asymmetry, a behavior also observed in the Kerr case. Thus, the parameters $\xi$ and $a$ exert competing influences on the shadow's morphology, i.e., the deformation parameter $\xi$ dominates the overall scale and circularity, deviating from the Kerr prediction, while the rotation parameter $a$ governs the degree of rotational distortion.

On the other hand, the angular diameter $\mathcal{D} = 2 \tilde{R}_d \frac{\mathcal{M}}{D_o}$ is used in astronomical observations to describe the size of the black hole shadow \cite{Amarilla:2011fx,Li:2024ctu}. Here, $D_o$ denotes the distance between the black hole and the observer, $\tilde{R}_d$ is the shadow radius when the screen is placed at the black hole's location, which is related to the shadow size $R_d$ and can be calculated using a simple geometric relation, and $\mathcal{M}$ is the mass of the black hole. When the black hole is far from the observer, the angular diameter can be quantitatively expressed as
\begin{equation}
	\mathcal{D} = 2 \times 9.87098\tilde{R}_{d}\left(\frac{\mathcal{M}}{M_{\odot}}\right)\left(\frac{1\mathrm{kpc}}{D_{o}}\right)\mathrm{\mu as}.\label{ar}
\end{equation}
Using the above equation, we calculate the theoretical angular diameters of Sgr A* and M87* under different parameters by assuming that their background spacetime is given by equation~(\ref{metric1}), and compare the results with astronomical observations. For M87*, its distance from Earth is $D_o = 16.8\ \mathrm{Mpc}$, and its estimated black hole mass is $\mathcal{M} = (6.5 \pm 0.7) \times 10^6 M_\odot$, while the observed shadow diameter is $\mathcal{D}_{\mathrm{M87^*}} = (37.8 \pm 2.7)\ \mathrm{\mu as}$ \cite{Capozziello:2023tbo}. For Sgr A*, its distance from Earth is $D_o = 8\ \mathrm{kpc}$, and its estimated black hole mass is $\mathcal{M} = \left(4.0_{-0.6}^{+1.1}\right) \times 10^6 M_\odot$, while the observed shadow diameter is $\mathcal{D}_{\mathrm{Sgr A^*}} = \left(48.7 \pm 7\right)\ \mathrm{\mu as}$ \cite{KumarWalia:2022aop}.

\begin{figure*}[htbp]
	\centering
	\subfigure[$a=0.5$]{\includegraphics[width=0.35\textwidth]{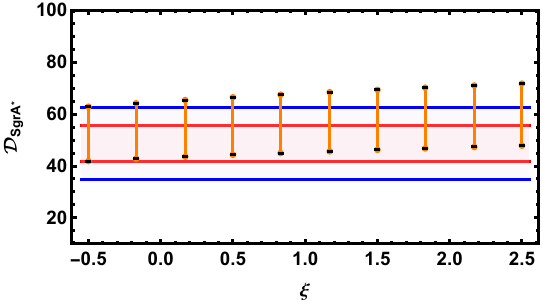}}
	\subfigure[$a=1.15$]{\includegraphics[width=0.35\textwidth]{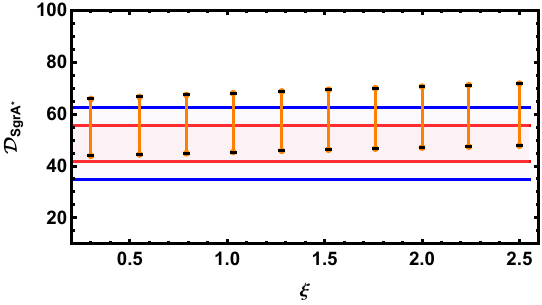}}
	\subfigure[$a=0.5$]{\includegraphics[width=0.35\textwidth]{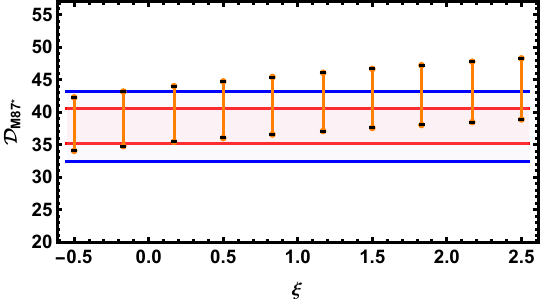}}
	\subfigure[$a=1.15$]{\includegraphics[width=0.35\textwidth]{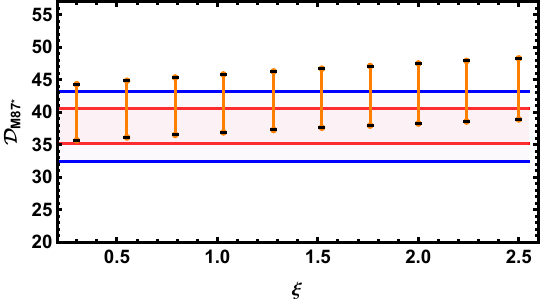}}
	\caption{\label{CI} Estimated intervals of the angular diameter $\mathcal{D}$. The first row corresponds to Sgr A*, and the second row corresponds to M87*. In the figure, the red and blue solid lines denote the $1\sigma$ and $2\sigma$ confidence intervals of $\mathcal{D}$, respectively, while the orange line segments represent the estimated intervals, with their endpoints highlighted by thick black ticks.}
\end{figure*}

In Figure~\ref{CI}, we present the estimated intervals of the angular diameters $\mathcal{D}$ for Sgr A* and M87*. The first row corresponds to Sgr A*, and the second row corresponds to M87*. In the figure, the red solid lines represent the $1\sigma$ confidence interval of $\mathcal{D}$, the blue solid lines represent the $2\sigma$ confidence interval, and the orange line segments indicate the estimated intervals, with their endpoints highlighted by thick black ticks. When $M > a = 0.5$, for $-0.5 \leq \xi \leq 2.5$, the estimated intervals of the angular diameters for both Sgr A* and M87* increase with $\xi$. Similarly, when $M < a = 1.15$, the estimated intervals of the angular diameters for both objects also increase with $\xi$. All the estimated intervals are within the $1\sigma$ confidence interval, indicating that whether $a < M$ or $a > M$, the astronomical observations impose weak constraints on the deformation parameter $\xi$ for both M87* and Sgr A*.

\section{Thin accretion disk model  and the optical appearance of black hole }
\label{sec4}
Considering that the millimeter-wave images of supermassive black holes are predominantly influenced by the presence of their accretion disks, we will employ a specific model that represents the accretion disk as an illuminating source to further investigate the imaging characteristics of a Konoplya-Zhidenko rotating non-Kerr black hole. The aim  is to clarify the distinct impact on the observable characteristics of accretion disk structures that arises from modifications to the geometric properties of spacetime within a Konoplya-Zhidenko rotating non-Kerr black hole paradigm.

\subsection{The selection of the observer}
\begin{figure}[!h]
\makeatletter
\renewcommand{\@thesubfigure}{\hskip\subfiglabelskip}
\makeatother
\centering 
\subfigure[]{
\setcounter{subfigure}{0}\subfigure[]{\includegraphics[width=0.45\textwidth]{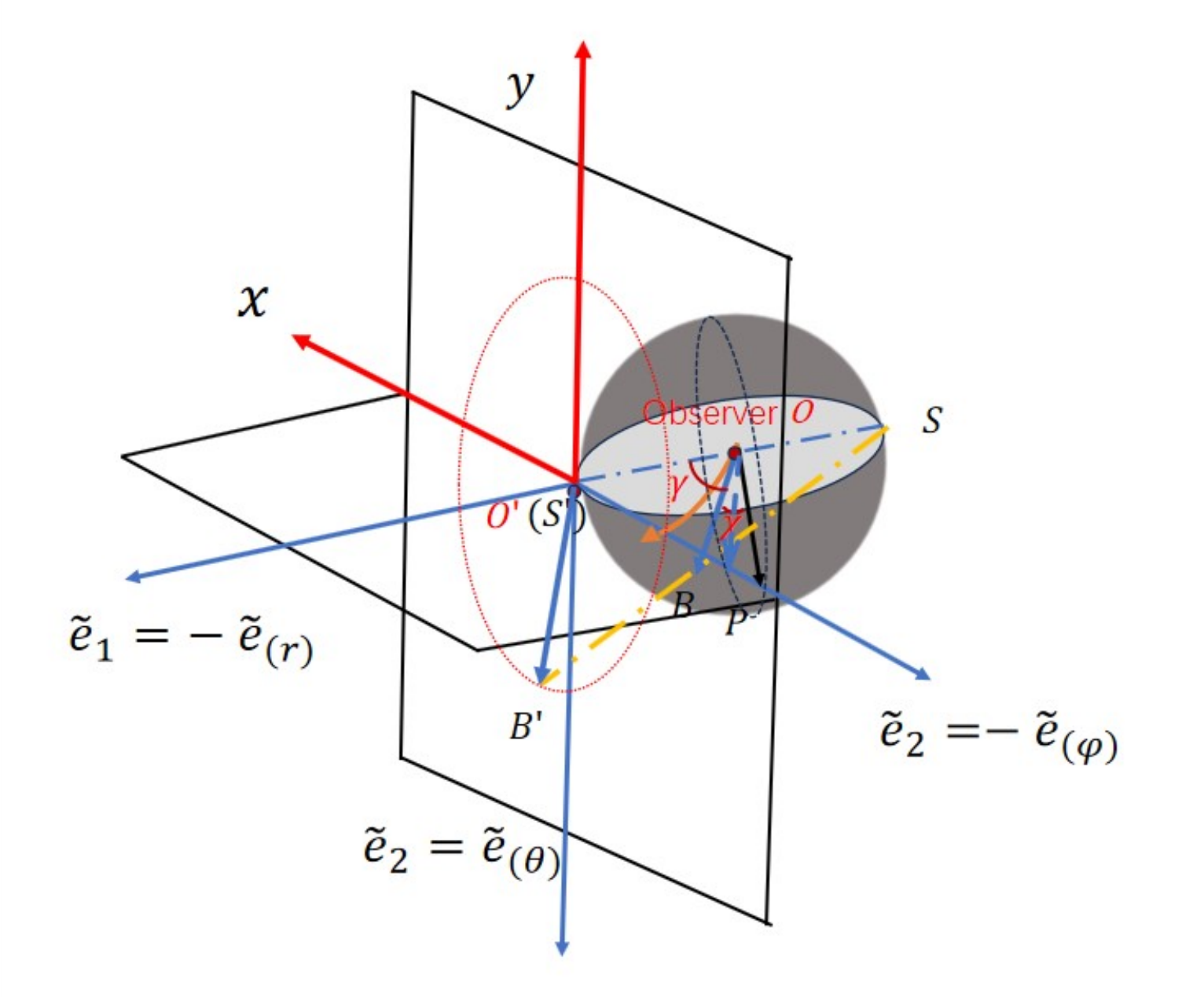}}}
\caption{\label{figZ} The ZAMO tetrad and celestial coordinates $(\gamma, \chi)$ based on the method of stereographic projection. The gray sphere situated on the right side of the image symbolizes the three-dimensional subspace from the observer's perspective, whereas the plane $(x, y)$ denotes the imaging plane. }
\end{figure}

As a general selection, we consider an observer  with zero axial angular momentum, also known as a zero angular momentum observer (ZAMO).
When the  position of ZAMO  is denoted as $(t_O, r_O, \theta_O, \varphi_O)$ in the spacetime, a locally orthonormal frame can be established within the vicinity of the observer, which is
\begin{align} \label{OT1}
\tilde{e}_0=\tilde{e}_{(t)}=\left(\sqrt{\frac{-g_{\varphi \varphi}}{g_{tt}g_{\varphi\varphi}-g^2_{t \varphi}}}, 0, 0, -\frac{g_{t \varphi}}{g_{\varphi \varphi}}\sqrt{\frac{-g_{\varphi \varphi }}{g_{tt}g_{\varphi \varphi }-g^2_{t \varphi}}}\right), \quad \tilde{e}_1=-\tilde{e}_{(r)}=\left(0, -\frac{1}{\sqrt{g_{rr}}}, 0, 0\right),
\end{align}
\begin{align} \label{OT3}
\tilde{e}_2=\tilde{e}_{(\theta)}=\left(0, 0, \frac{1}{\sqrt{g_{ \theta  \theta}}},  0\right), \quad \tilde{e}_3=-\tilde{e}_{(\varphi)}=\left(0, 0, 0, -\frac{1}{\sqrt{g_{ \varphi \varphi}}}\right).
\end{align}
Here, the timelike vector $\tilde{e}_0$ is considered as the observer's four-velocity,  the vector $\tilde{e}_1$ represents the spatial direction towards the center of the black hole, while $g_{\mu\nu}$ is the component of the background metric at $(r_O, \theta_O)$. After determining the full light trajectory, the next step is to consider how to image the black hole. One direct approach is to employ a pinhole camera for perspective projection, as described in \cite{Wang:2017hjl}. This model is straightforward and aligns well with the actual imaging principles; however, it has the limitation of a relatively narrow field of view. Therefore, we will utilize the imaging methodology detailed in \cite{Hu:2020usx}, commonly known as the fisheye lens camera model. By employing the stereographic projection technique, the detection of a photon in the image plane corresponds to the optical perspective of an observer, who can be considered analogous to a camera, see Figure \ref{figZ}.

In Figure \ref{figZ}, the observer's position within the coordinate system is denoted by point $O$, while the orange curve with an arrow originating from point $O$ represents the direction of light propagation. The vector $\overrightarrow{OB}$ denotes the tangential vector of the null geodesic at $O$. To further derive the black hole shadow and its corresponding image, one can construct a three-dimensional sphere with the $|\overrightarrow{OB}|$ as its radius and $O$ as the center, positioning the origin of the tetrad (\ref {OT1}) at point $S'$. The diameter $SS'$ of this three-dimensional sphere is on the same straight line as $\tilde{e}_1$. The rectangular plane positioned at point $S'$ serves as the screen for displaying the image of the black hole shadow. Therefore, the vector $\overrightarrow{OB}$  is projected onto the screen as the vector $\overrightarrow{O'B'}$.
It is useful to introduce the celestial sphere coordinates $(\gamma, \chi)$ in order to determine the position of a photon from the observer's perspective.  The vertical plane of the sphere intersects the plane ($O'BS$) at point $P$, and another celestial coordinate  $\chi$ is defined as the angle between $OP$ and  $\tilde{e}_2$. By employing celestial coordinates $(\gamma, \chi)$, the tangent vectors of null geodesics followed by photons, i.e., $s(\tau)=\{t(\tau), r(\tau), \theta(\tau), \varphi(\tau)\}$, in the observer's frame can be expressed as linear combinations of $(\tilde{e}_0, \tilde{e}_1, \tilde{e}_2, \tilde{e}_3)$, that is
\begin{align} \label{TV1}
\dot{s}=|\overrightarrow{OB}|(-\delta \tilde{e}_0+ \cos \gamma \tilde{e}_1+\sin\chi\cos\gamma \tilde{e}_2+ \sin\gamma\sin\chi \tilde{e}_3 ).
\end{align}
In the above equation, the negative sign before $\tilde{e}_0$ is introduced to ensure that the tangent vector points forward, while the symbol $\cdot$ denotes the partial derivative with respect to the affine parameter $\tau$. Since the trajectory of a photon is independent of its energy, for simplicity, one  can set the energy of the photon observed by the camera to $1$, which is
\begin{align} \label{EN1}
E_C=1= |\overrightarrow{OB}|\cdot \delta=-\frac{E}{\sqrt{g_{tt}}}\mid_{(r_O, \theta_O)}.
\end{align}
In the frame of ZAMO, the four-momentum of photons  is denoted as $p_{(\mu)}=p_\nu \tilde{e}^\nu_{(\mu)}$, and the term of $p_{(\mu)}$ is derived from equations (\ref {GE1})-(\ref{GE4}), while  $\tilde{e}^\nu_{(\mu)}$ can be obtained through equations (\ref {OT1})-(\ref{OT3}). The relationship between the four-momentum of a photon $p_{(\mu)}$ and celestial coordinates $(\gamma, \chi)$ is established with
\begin{align} \label{TM3}
\cos \gamma=\frac{p^{(1)}}{p^{(0)}}, \qquad \tan \chi= \frac{p^{(3)}}{p^{(2)}}.
\end{align}
The acquisition of an image of a black hole requires the transformation of celestial coordinates $(\gamma, \chi)$ onto the observation plane.
The imaging plane can be equipped with a standard Cartesian coordinate system $(x, y)$, which establishes a precise correspondence with the celestial coordinates, which is
\begin{align} \label{TM3}
x=-2 |\overrightarrow{OB}| \tan\frac{\gamma}{2}\sin \chi, \qquad y= -2 |\overrightarrow{OB}| \tan \frac{\gamma}{2}\cos \chi.
\end{align}
In essence, it determines the initial momentum value of the photon at the ZAMO as well as its initial position.
To analyze the shape and size of the black hole shadow seen by equatorial observers under a region of strong gravitational field, one can map the shadow image of a Konoplya-Zhidenko rotating non-Kerr black hole at different $\xi$ and $a$ values.

\subsection{Configuration of the thin accretion disk model}
\begin{figure}[!h]
\makeatletter
\renewcommand{\@thesubfigure}{\hskip\subfiglabelskip}
\makeatother
\centering 
\subfigure[]{
\setcounter{subfigure}{0}\subfigure[]{\includegraphics[width=0.45\textwidth]{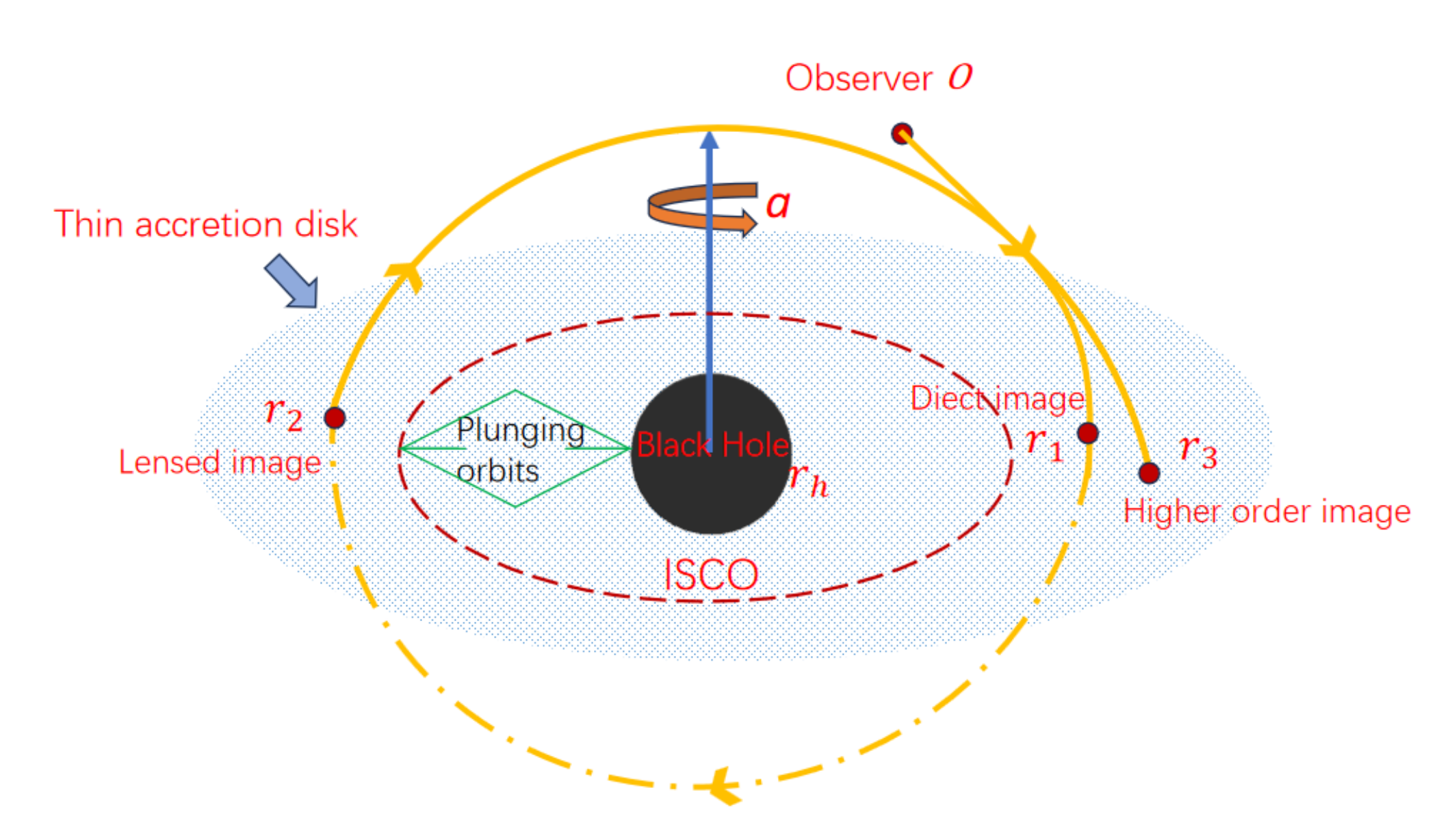}}}
\caption{\label{figob1} Imaging the black hole with a thin accretion disk, where the black sphere denotes the black hole, the blue elliptical disk represents the thin accretion disk, and the red dashed line indicates the ISCO. Additionally, the orange curve represents a complete path of light that can be received by the observer.}
\end{figure}

The analysis of the accretion disk can offer valuable insights into the fundamental physical mechanisms governing this intricate system. Therefore, it is necessary to elucidate certain specific aspects of this accretion model: i) the accretion disk model is characterized by its extreme thinness in the geometric and optical, being situated precisely on the equatorial plane; ii) the material comprising the accretion disk consists of electrically neutral free plasma that moves along an equatorial timelike geodesic; iii) in this work, only black hole configurations admitting an event horizon are considered. Accordingly, the thin accretion disk is assumed to extend from the event horizon $r_h$ to a radius beyond the innermost stable circular orbit (ISCO). The ISCO is an important indicator, and its radius can be used along with the radius of the inner boundary of the accretion disk to calculate the energy emission efficiency, which measures how efficiently the rest mass energy of matter is converted into radiative energy. Moreover, the viscosity of the accretion disk  induces an outward transfer of angular momentum by the orbiting accretion material around the black hole, resulting in a gradual migration of particles towards the ISCO region in Keplerian motion. The particles undergo acceleration and spiral towards the event horizon until they reach the black hole, and it has been experimentally confirmed that the particle motion mechanism during this process is consistent with astrophysical observations\cite{Chael:2021rjo}. In recent work \cite{Hou:2022eev}, the behavior of particle motion in the accretion disk has been well explained.  Specifically, the ISCO serves as a demarcation line that segregates particle behavior within the accretion disk into two distinct categories: the region encompassing ISCO ($r<r_{isco}$), where particles undergo a critical plunging orbit, and the outer region beyond ISCO ($r>r_{isco}$), where particles maintain a stable circular orbit. In the spacetime of (\ref{metric1}), the ISCO position of a black hole is given by
\begin{align} \label{VEFF}
V_{e}(r)|_{r=r_{isco}}=0, \qquad  \partial_r V_{e}(r)|_{r=r_{isco}}=0, \qquad \partial^2_r V_{e}(r)|_{r=r_{isco}}=0.
\end{align}
The term $V_{e}(r)$  denotes the effective potential, and a massive neutral particle in the equatorial plane $(\theta = \pi/2)$  with four-velocity $u^a$ can be defined as
\begin{align} \label{VEFF2}
V_{e}(r, \mathcal{E}, \mathfrak{L})=1+g^{tt}\mathcal{E}^2+g^{tt}\mathfrak{L}^2-2g^{t\varphi}\mathcal{E}\mathfrak{L}.
\end{align}
Here, $\mathcal{E}$ and $\mathfrak{L}$ are two conserved quantities, denoting the specific energy and specific angular momentum of the massive neutral particle, that is
\begin{align} \label{EN1}
\mathcal{E}=-\frac{g_{tt}+g_{t\varphi}\overline{\mathbf{W}}}{\sqrt{-g_{tt}-2g_{t\varphi}\overline{\mathbf{W}}-g_{\varphi \varphi}\overline{\mathbf{W}}^2}}, \quad \mathfrak{L}=-\frac{g_{t\varphi}+g_{\varphi \varphi}\overline{\mathbf{W}}}{\sqrt{-g_{tt}-2g_{t\varphi}\overline{\mathbf{W}}-g_{\varphi \varphi}\overline{\mathbf{W}}^2}}.
\end{align}
In which,  $\overline{\mathbf{W}}$ denotes the angular velocity, which is
\begin{align} \label{WN1}
\overline{\mathbf{W}}=\frac{d\varphi}{dt}=\frac{\partial_rg_{t\varphi}+\sqrt{\partial^2_r g_{t\varphi}-\partial_r g_{tt}\partial_r g_{\varphi \varphi}}}{\partial_r  g_{\varphi \varphi}}
\end{align}
In the accretion disk within the range $r_h< r \leq  r_{isco}$, the particle falls towards the black hole on a critical plunging orbit, and its  four-velocity is
\begin{align} \label{UT1}
u^t_f=-g^{tt}\mathcal{E}_i+g^{t\varphi}\mathfrak{L}_i,
\end{align}
\begin{align} \label{UF1}
u^\varphi_f=-g^{t\varphi}\mathcal{E}_i+g^{\varphi \varphi}\mathfrak{L}_i,
\end{align}
\begin{align} \label{UC1}
u^\theta_f=0,
\end{align}
\begin{align} \label{UR1}
u^r_f=-\sqrt{-\frac{1+g_{tt}u^t_f u^t_f+2  g_{t \varphi} u^t_f u^\varphi_f+g_{\varphi \varphi}u^\varphi_f u^\varphi_f}{g_{rr}}}.
\end{align}
In the above equations, $\mathcal{E}_i$ and $\mathfrak{L}_i$ are the conserved quantities of the particle at ISCO, and the minus sign before the square root indicates that the direction is toward the black hole.
Outside the ISCO $r > r_{isco}$, the four-velocity $u_c^\mu$ is governed by
\begin{align} \label{UC2}
u_c^\mu= \sqrt{\frac{1}{-g_{tt}-2g_{t\varphi}\overline{\mathbf{W}}-g_{\varphi \varphi}\overline{\mathbf{W}}^2}}(1,0,0,\overline{\mathbf{W}}).
\end{align}
In the process of  tracing the light ray path, multiple intersections with the accretion disk in the equatorial plane may occur, resulting in varying radial coordinates at each intersection, see Figure \ref{figob1}.
Figure \ref{figob1} illustrates a complete photon trajectory (orange curve) approaching the observer. The black sphere denotes the black hole, the red dashed circle marks the ISCO, and the blue disk represents the thin accretion disk in the equatorial plane. Each time this light ray intersects the accretion disk at a radius $r_n$ ($n=1,2,...,N$), it contributes to the observed intensity. The transfer function $r_n(x,y)$ maps the screen coordinates to the $n$-th intersection point, generating the $n$-th order image: the direct image ($n=1$), the lensing ring ($n=2$), and the photon ring ($n\geq3$). Accordingly, the total observed intensity accumulates with each disk crossing. In this case, the intensity of the observation on the observation screen can be expressed as
\begin{align} \label{IO1}
\mathcal{I}_{o}=\sum^{N}_{n=1}f_n g_n^3 \mathcal{J}_n (r).
\end{align}
where $f_n$ is the fudge factor which is fixed to $f_n=1$. Taking into account that the black hole image captured by the EHT is observed at a wavelength of 1.3 mm (230 GHz), one can select the emissivity of the thin disk to be a  second-order polynomial in log-space, which is
\begin{align} \label{EM1}
\mathcal{J}(r)=\exp \left[-\frac{1}{2}\mathcal{Z}^2-2 \mathcal{Z}\right], \qquad \mathcal{Z}=\log\frac{r}{r_h}.
\end{align}
In addition, the term $g_n$ represents the redshift factor, which is defined as the ratio of the observed frequency to the emission frequency at radius $r_n$, that is
\begin{align} \label{RF}
g_n=\frac{\nu_{obs}}{\nu_n}.
\end{align}
Here, the observed frequency by the observer is denoted as $\nu_{obs}$, while the frequency observed by the local static frames co-moving with the emission profiles is represented as $\nu_n$. It is important to note that in equation (\ref{IO1}), the third power of the redshift factor $g_n$ is employed, as it corresponds to the intensity at the specific frequency of 230 GHz \cite{Wang:2023fge}.  In the work of Gralla\cite{Gralla:2020srx}, the application of the fourth power form of the redshift factor $g_n$ was also elaborately introduced. Naturally, the specific forms of the redshift factor differ in the inner and outer regions of the ISCO due to significant disparities in particle emission spectra between these two domains. The accretion flow beyond the ISCO moves along a circular orbit, and its corresponding redshift factor $g^{out}_{n}$ can be expressed as
\begin{align} \label{RF2}
g^{out}_{n}=\frac{\eta(1-\upsilon \frac{p_\varphi}{p_t})}{\zeta(1+\overline{\mathbf{W}}\frac{p_\varphi}{p_t})}\mid_{r=r_n}, \qquad r > r_i,
\end{align}
where
\begin{align} \label{RF3}
\upsilon=\frac{g_{t \varphi}}{g_{\varphi \varphi}}, \qquad \eta=\sqrt{-\frac{g_{\varphi \varphi}}{g_{tt}g_{\varphi \varphi}-g^2_{t \varphi}}}, \qquad
\zeta=\sqrt{\frac{-1}{g_{tt}+2g_{t \varphi}\overline{\mathbf{W}}+g_{\varphi \varphi}\overline{\mathbf{W}}^2}}.
\end{align}
In addition, the ratio of the observed energy on the screen to the energy along a null geodesic is $e=\frac{p_{(t)}}{p_t}$, that is
\begin{align} \label{RF4}
e=\frac{p_{(t)}}{p_t}=\eta(1-\upsilon \frac{p_\varphi}{p_t}).
\end{align}
In the context of asymptotically flat spacetime, when the observer is positioned at an infinite distance, it is permissible to assign a value of $e$ as $e=1$. In the region within ISCO, the accretion flow is moving along the critical plunge orbit, and its redshift factor $g^{in}_{n}$ is
\begin{align} \label{RFP}
g^{in}_{n}=-\frac{1}{u^r_f p_r/p_t-\mathcal{E}_i(g^{tt}-g^{t\varphi} p_\varphi/p_t)+\mathfrak{L}_i(g^{\varphi \varphi}p_\varphi/p_t+g^{t\varphi})}\mid_{r=r_n}, \qquad r < r_i.
\end{align}
Within the framework of the accretion thin disk model, one can explicitly simulate the visual representation of a Konoplya-Zhidenko rotating non-Kerr black hole on the display by using the redshift factor and the emission model of the disk.

\subsection{Observational appearance of the black hole}
The visualization image of the Konoplya-Zhidenko rotating non-Kerr black hole illuminated by the accretion disk can be achieved using the backward ray-tracing technique, which involves tracing the light path backwards from the observer's position, based on the aforementioned accretion thin disk model.  The inner radius of the accretion disk is set to match the event horizon $r_{in}=r_h$, while the outer radius is defined as $r_{out} = 20M$, thus achieving an extension of the accretion disk up to the event horizon. By considering the motion of the accretion disk, one can account for two possibilities: prograde and retrograde, acknowledging the presence of both forward and backward photons in the Konoplya-Zhidenko rotating non-Kerr black hole.

\begin{figure}[!h]
\centering 
\subfigure[$a=0.3$, $\theta_o = 0^{\circ}$, $\xi=-0.9$]{\includegraphics[scale=0.5]{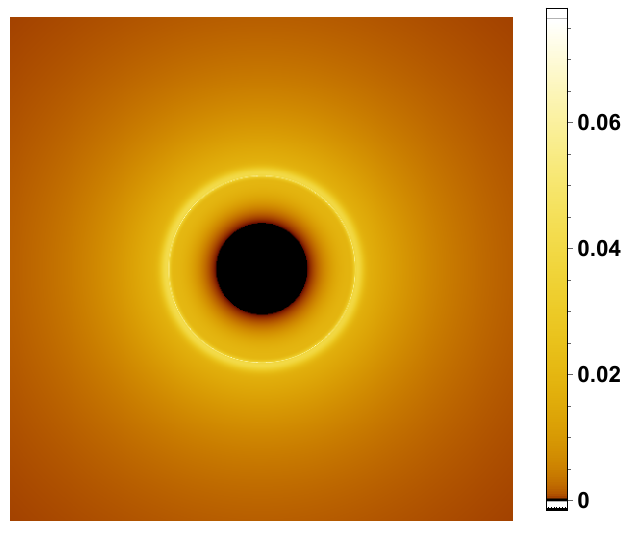}}
\subfigure[$a=0.3$, $\theta_o = 17^{\circ}$, $\xi=-0.9$]{\includegraphics[scale=0.5]{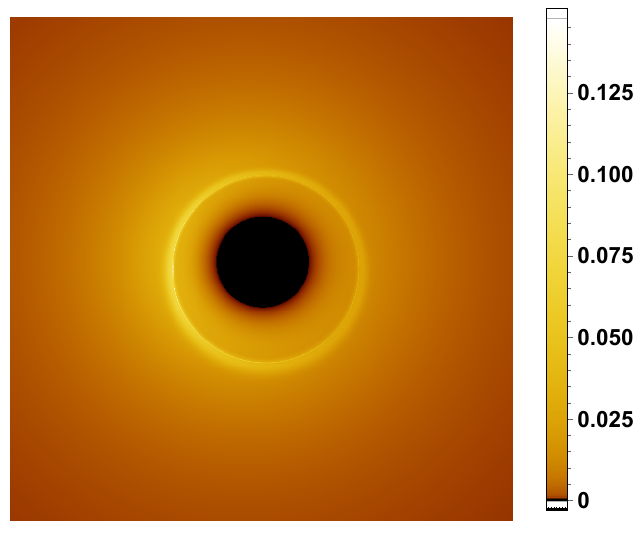}}
\subfigure[$a=0.3$, $\theta_o = 83^{\circ}$, $\xi=-0.9$]{\includegraphics[scale=0.5]{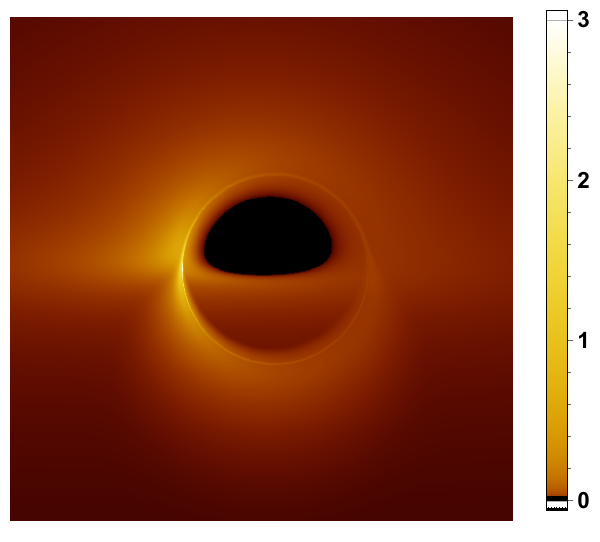}}
\subfigure[$a=0.3$, $\theta_o = 0^{\circ}$, $\xi=0.9$]{\includegraphics[scale=0.5]{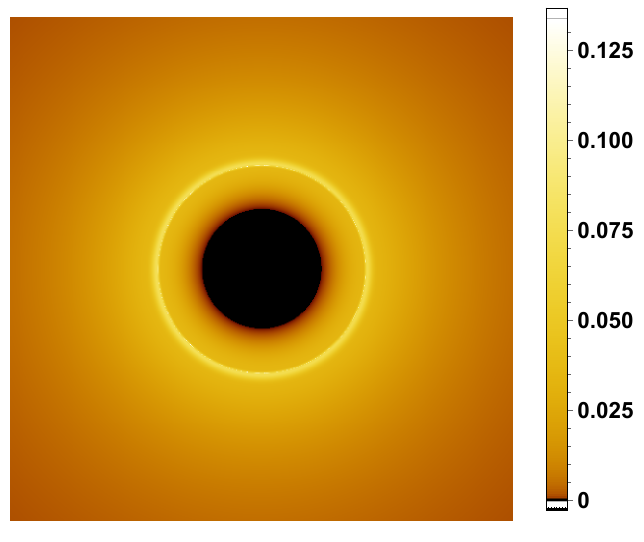}}
\subfigure[$a=0.3$, $\theta_o = 17^{\circ}$, $\xi=0.9$]{\includegraphics[scale=0.5]{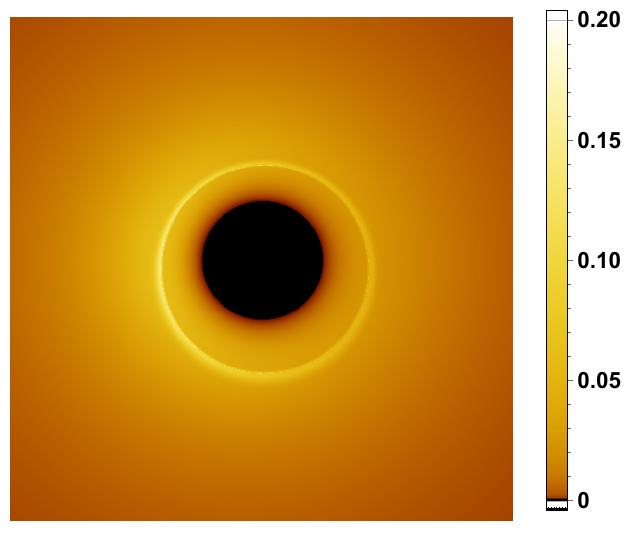}}
\subfigure[$a=0.3$, $\theta_o = 83^{\circ}$, $\xi=0.9$]{\includegraphics[scale=0.5]{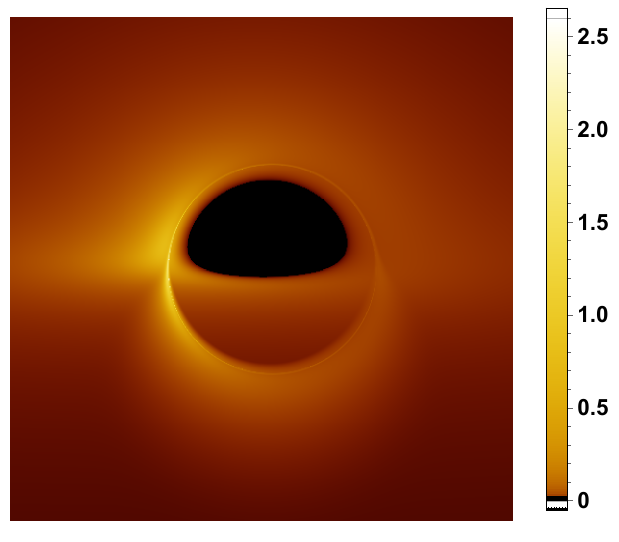}}
\subfigure[$a=0.99$, $\theta_o = 0^{\circ}$, $\xi=0.9$]{\includegraphics[scale=0.5]{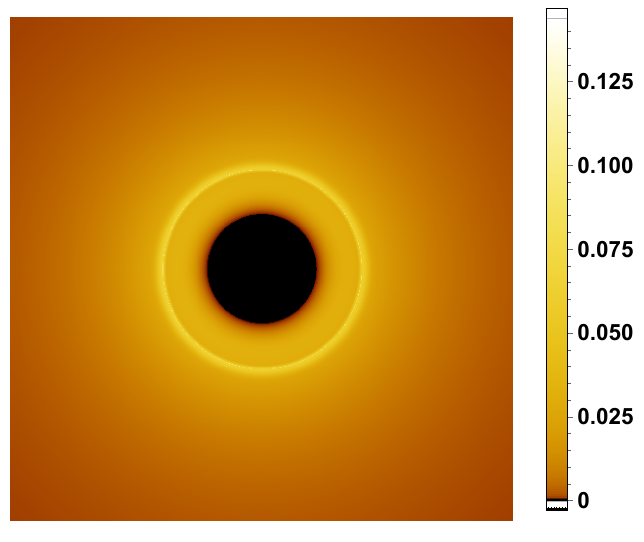}}
\subfigure[$a=0.99$, $\theta_o = 17^{\circ}$, $\xi=0.9$]{\includegraphics[scale=0.5]{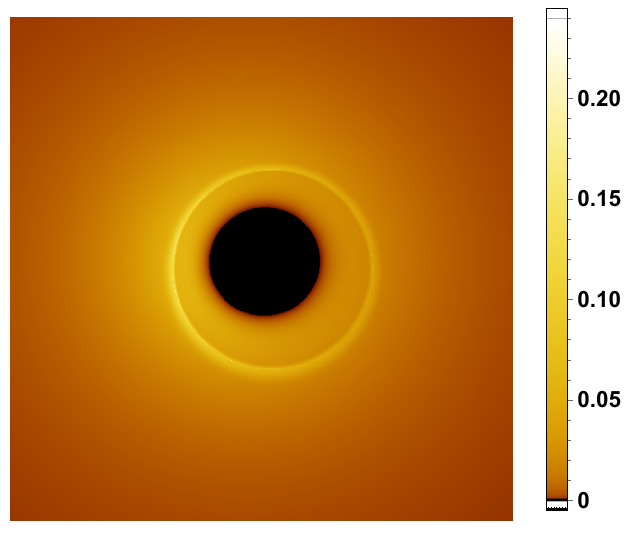}}
\subfigure[$a=0.99$, $\theta_o = 83^{\circ}$, $\xi=0.9$]{\includegraphics[scale=0.5]{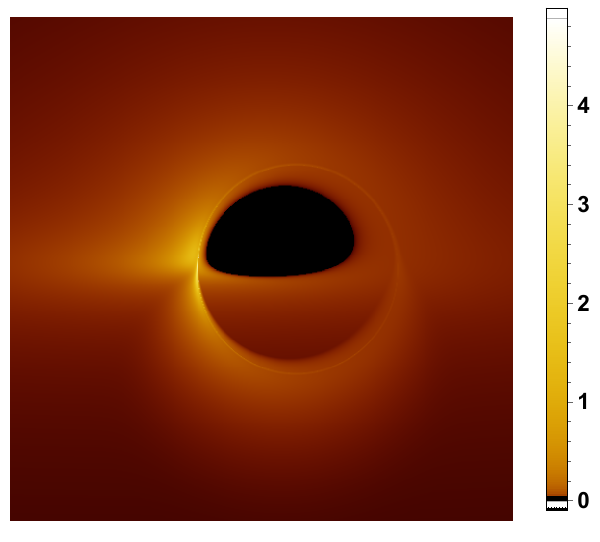}}
\caption{\label{figbpA1}  The image of the Konoplya-Zhidenko rotating non-Kerr black hole surrounded by a prograde thin accretion disk at 230 GHz,
 where the parameter values are taken as $a<M$. For all images, the mass $ M $ is set to $ M = 1$, the observer's position is fixed at $ r_{obs} = 500M $, and the black hole's event horizon is depicted as a black region. }
\end{figure}

\begin{figure}[!h]
\centering 
\subfigure[$a=1.05$, $\theta_o = 0^{\circ}$, $\xi=0.5$]{\includegraphics[scale=0.5]{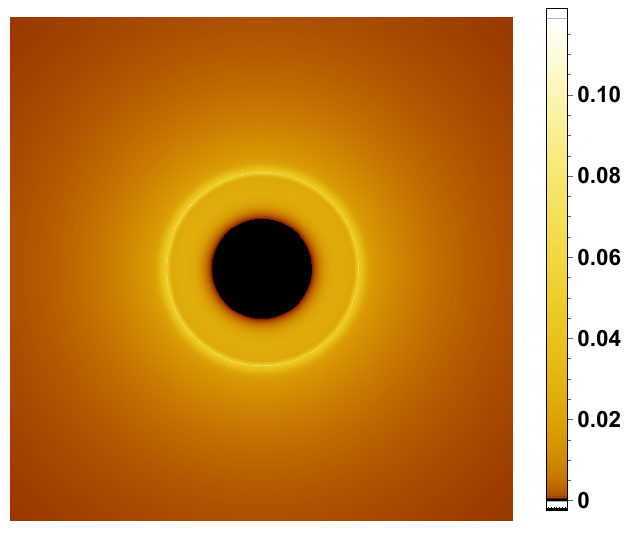}}
\subfigure[$a=1.05$, $\theta_o = 17^{\circ}$, $\xi=0.5$]{\includegraphics[scale=0.5]{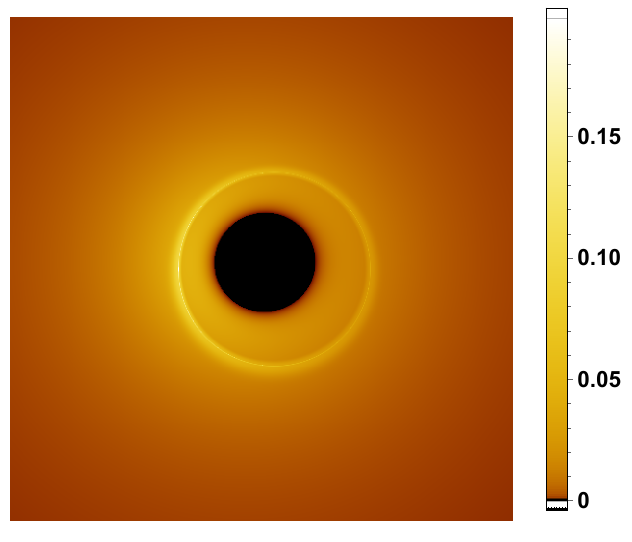}}
\subfigure[$a=1.05$, $\theta_o = 83^{\circ}$, $\xi=0.5$]{\includegraphics[scale=0.5]{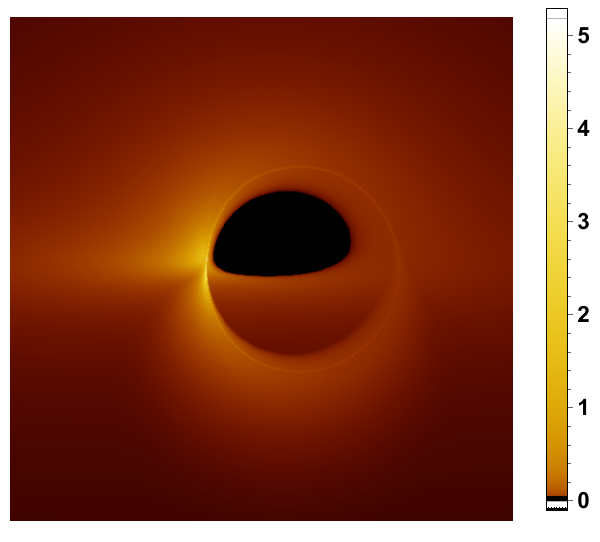}}
\subfigure[$a=1.05$, $\theta_o = 0^{\circ}$, $\xi=2$]{\includegraphics[scale=0.5]{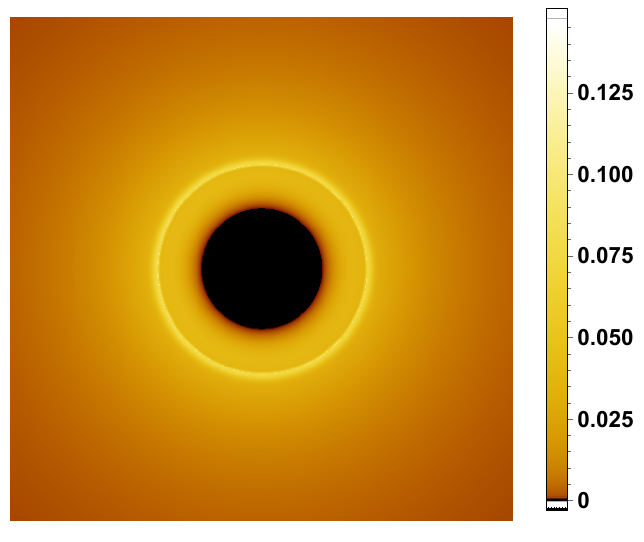}}
\subfigure[$a=1.05$, $\theta_o = 17^{\circ}$, $\xi=2$]{\includegraphics[scale=0.5]{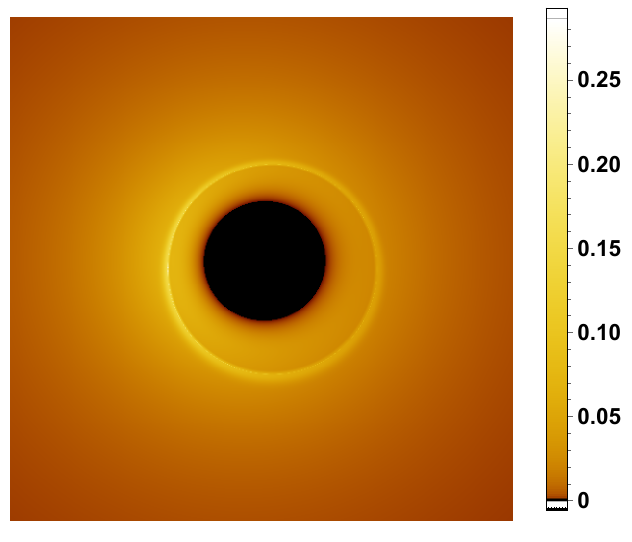}}
\subfigure[$a=1.05$, $\theta_o = 83^{\circ}$, $\xi=2$]{\includegraphics[scale=0.5]{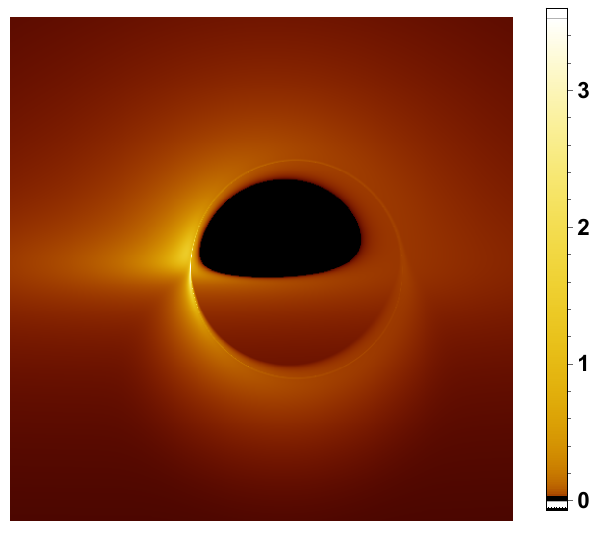}}
\subfigure[$a=1.15$, $\theta_o = 0^{\circ}$, $\xi=0.5$]{\includegraphics[scale=0.5]{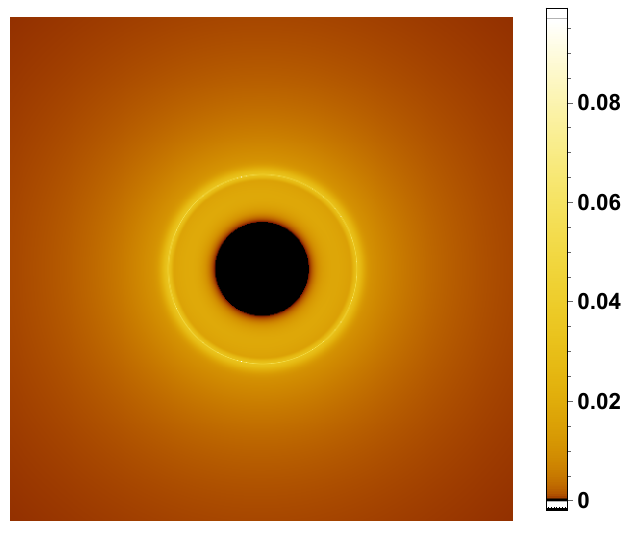}}
\subfigure[$a=1.15$, $\theta_o = 17^{\circ}$, $\xi=0.5$]{\includegraphics[scale=0.5]{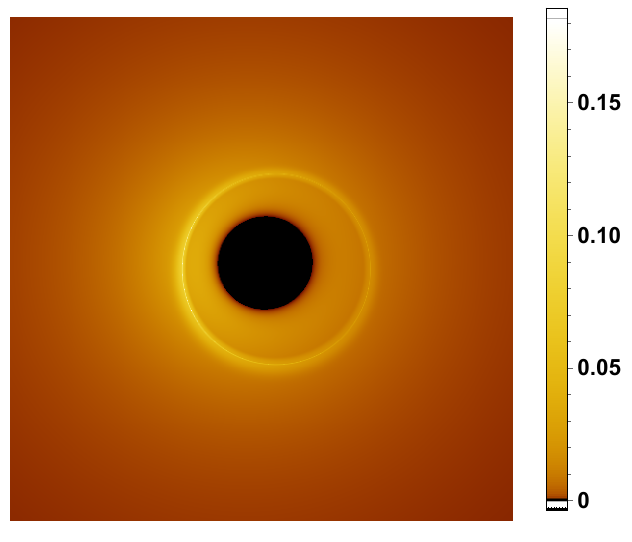}}
\subfigure[$a=1.15$, $\theta_o = 83^{\circ}$, $\xi=0.5$]{\includegraphics[scale=0.5]{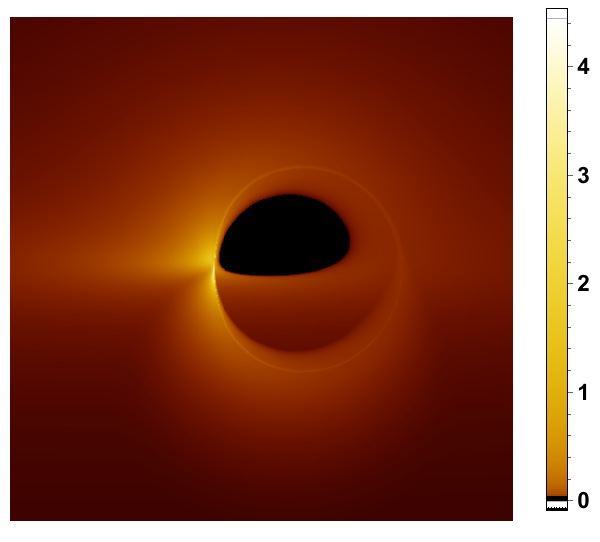}}
\caption{\label{figbpA2} The image of the Konoplya-Zhidenko rotating non-Kerr black hole surrounded by a prograde thin accretion disk at 230 GHz, where the parameter values are taken as $a>M$. }
\end{figure}

Under thin disk accretion with prograde flows, Figure \ref{figbpA1} and Figure \ref{figbpA2} illustrate the optical appearance of a Konoplya-Zhidenko rotating non-Kerr black hole with different parameter configurations. The selection of three distinct observation inclination angles includes one angle, $\theta_o=17^{\circ}$, which corresponds to the imaging angle for M87$^*$ as determined by the EHT. The case depicted in Figure \ref{figbpA1} corresponds to situations where the value of the rotation parameter $a$ is lower than $M$ $(a<M)$, whereas Figure  \ref{figbpA2} represents scenarios where the parameter $a$ exceeds that of the mass $M$ $(a>M)$.
It can be clearly observed that regardless of changes in the parameter value and observation angle, a black region always exists in the image. This is because some photons emitted within the event horizon are captured by it, while others manage to escape towards infinity. The photons that fail to reach the observer create a black region known as the inner shadow. However, the inner shadow region undergoes progressive deformation as the viewing angle increases, transitioning from a disk-like structure ($\theta_o=0^{\circ}$) to a hat-like shape ($\theta_o=83^{\circ}$). Another notable characteristic is that, regardless of variations in the system's parameters, a distinct bright closed-loop curve consistently encircles the outer boundary of the inner shadow region, marking the position of the critical curve. The significance lies in the fact that irrespective of the spin and deformation parameters of the black hole, the internal shadows and critical curves remain discernible under both low and high angle observations, thereby indicating their intrinsic nature as spacetime features of the black hole.

\begin{figure}[!h]
	\centering 
	\subfigure[$\theta_o = 0^{\circ}$]{\includegraphics[scale=0.5]{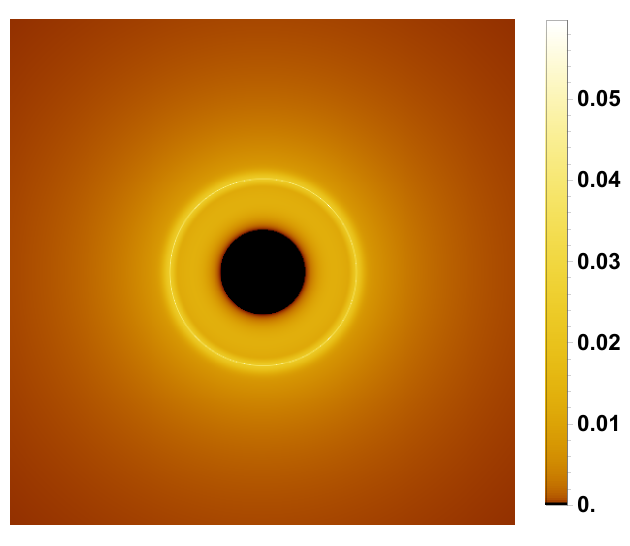}}
	\subfigure[$\theta_o = 17^{\circ}$]{\includegraphics[scale=0.5]{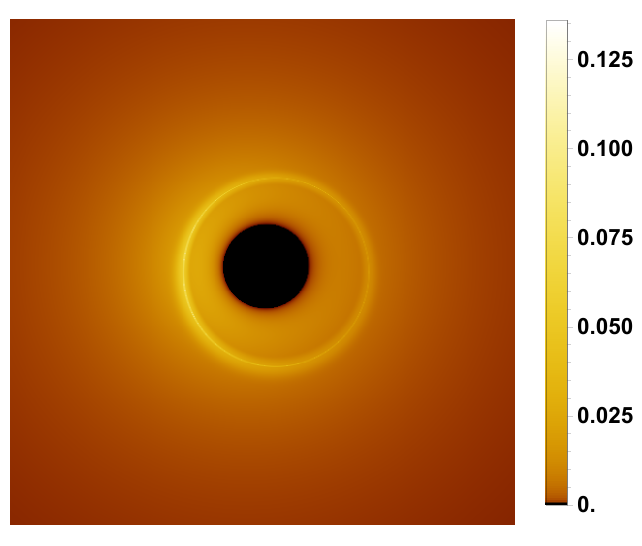}}
	\subfigure[$\theta_o = 83^{\circ}$]{\includegraphics[scale=0.5]{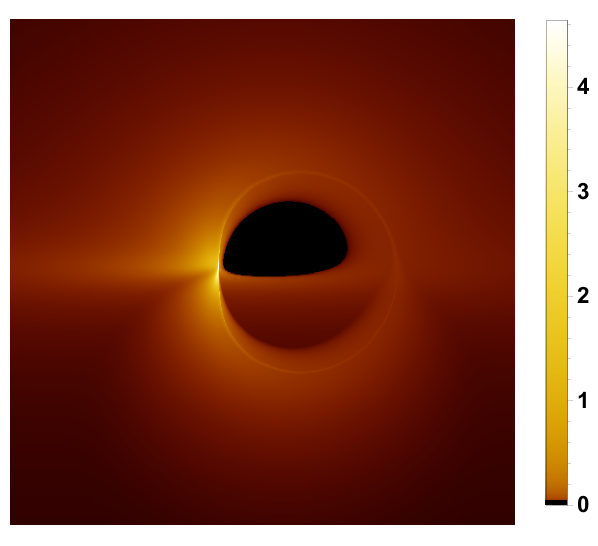}}
	\caption{\label{Kerr} The image of the Kerr black hole surrounded by a prograde thin accretion disk at 230 GHz, where the parameter values are taken as $a=0.99$.}
\end{figure}

In Figure \ref{figbpA1}, i.e., $a<M$, an increase in the deformation parameter $\xi$ will result in an enlargement of the  inner shadow region (first and second columns), whereas an increase in the rotation parameter $a$ will lead to a reduction of its size (second and third columns). Therefore, changes in rotation parameter $a$  and deformation parameters $\xi$ show mutually exclusive effects on the  size of inner shadow.
Additionally, the observed images of the bright bands of light surrounding the black hole at low observation angles, i.e., $\theta_o = 0^{\circ}$ (first line) and $\theta_o = 17^{\circ}$ (second line), reveal such a high concentration in brightness distribution that distinguishing between direct and lensed images becomes challenging. To put it more precisely, when the observational inclination is low, the secondary image of the black hole is incorporated into the primary image. When $\theta_o = 83^{\circ}$ (third line), the direct image and the lensed image of the black hole are distinguishable. The increase of the deformation parameter $\xi$ and rotation parameter $a$ in Figure \ref{figbpA2}, i.e., when $a>M$, has a similar impact on the size of the inner shadow region as the increase of $a<M$. Moreover, distinguishing between the direct image and  lensed image is only possible at higher observation angles. Figure \ref{Kerr} shows the shadow of a Kerr black hole with $a=0.99$, which is presented for comparison with the third row of Figure \ref{figbpA2} corresponding to a superspinning Konoplya-Zhidenko black hole with $a>M$. It can be seen that the inner shadow and the critical curve of the Kerr black hole are smaller in size. It should be noted that, according to \cite{Wang:2017hjl}, $\xi$ can take sufficiently small values, in which case the Konoplya-Zhidenko black hole shadow exhibits a cuspidal structure. In some general static and axisymmetric spacetimes, non-planar bound photon orbits exist independently of the integrability of photon motion \cite{Cunha:2017eoe}. For the Konoplya-Zhidenko black hole considered here, when $\xi$ is sufficiently small, photon trajectories become non-integrable, and thus such images are not generated in the present work. This scenario provides a potential direction for future investigation.

\begin{figure}[!h]
\centering 
\subfigure[$a=0.3$, $\theta_o = 0^{\circ}$, $\xi=-0.9$]{\includegraphics[scale=0.5]{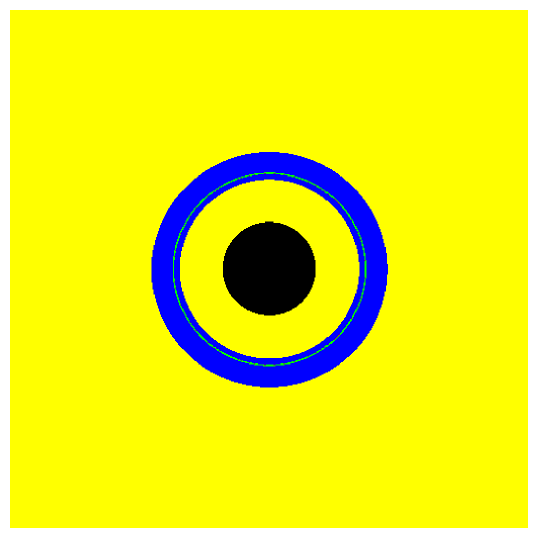}}
\subfigure[$a=0.3$, $\theta_o = 17^{\circ}$, $\xi=-0.9$]{\includegraphics[scale=0.5]{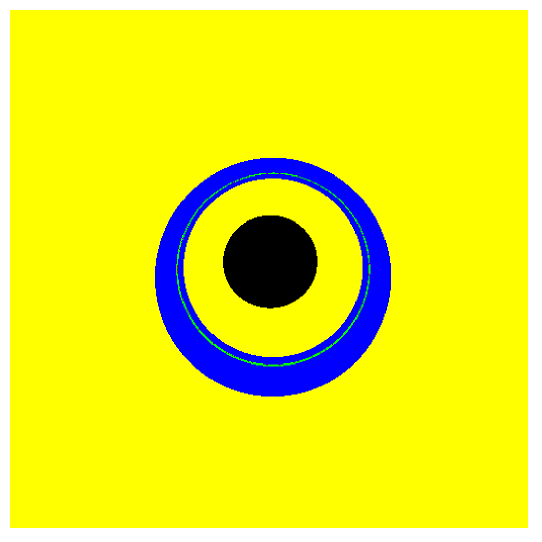}}
\subfigure[$a=0.3$, $\theta_o = 83^{\circ}$, $\xi=-0.9$]{\includegraphics[scale=0.5]{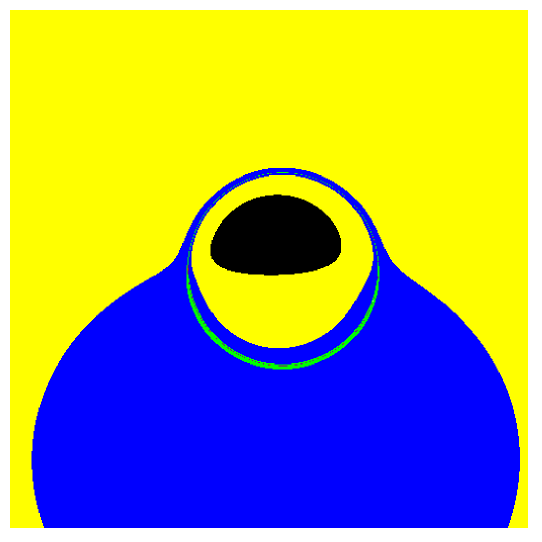}}
\subfigure[$a=0.3$, $\theta_o = 0^{\circ}$, $\xi=0.9$]{\includegraphics[scale=0.5]{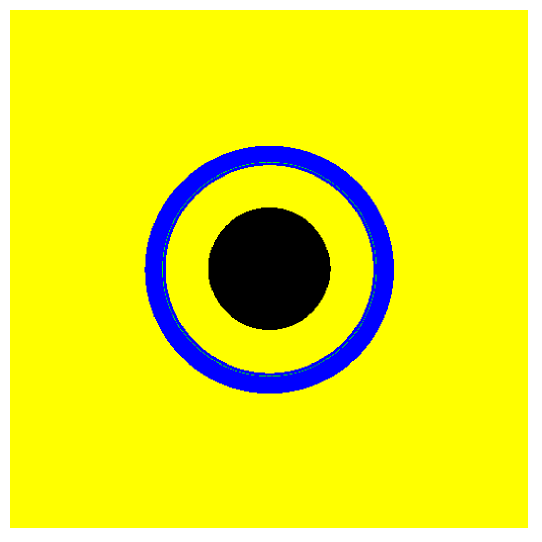}}
\subfigure[$a=0.3$, $\theta_o = 17^{\circ}$, $\xi=0.9$]{\includegraphics[scale=0.5]{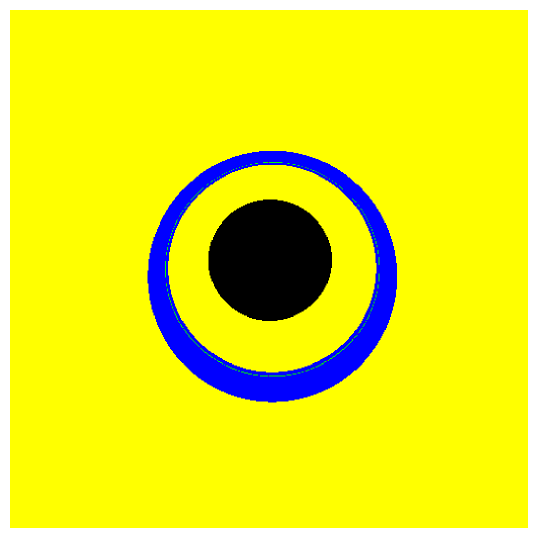}}
\subfigure[$a=0.3$, $\theta_o = 83^{\circ}$, $\xi=0.9$]{\includegraphics[scale=0.5]{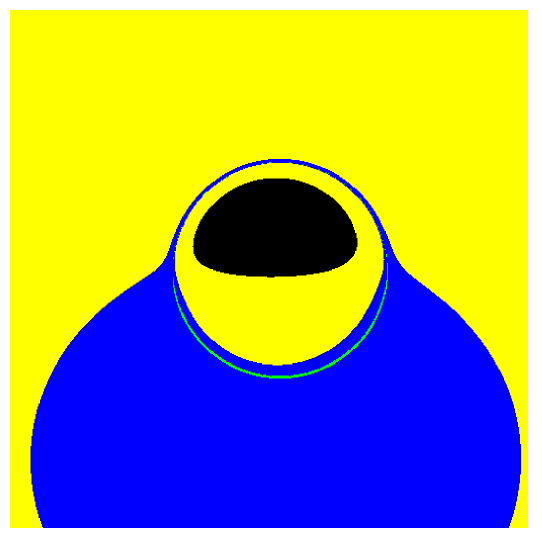}}
\subfigure[$a=0.99$, $\theta_o = 0^{\circ}$, $\xi=0.9$]{\includegraphics[scale=0.5]{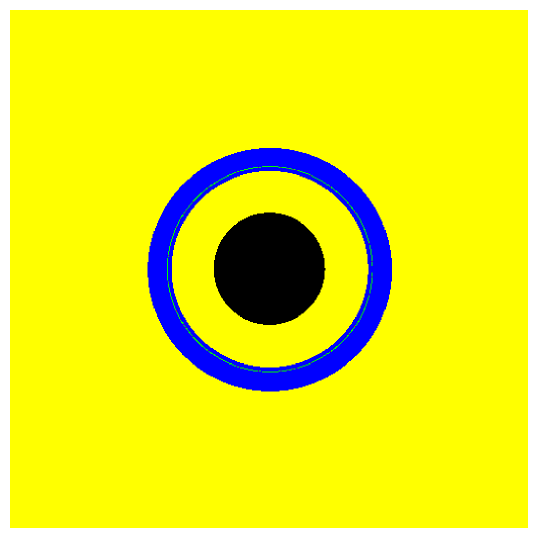}}
\subfigure[$a=0.99$, $\theta_o = 17^{\circ}$, $\xi=0.9$]{\includegraphics[scale=0.5]{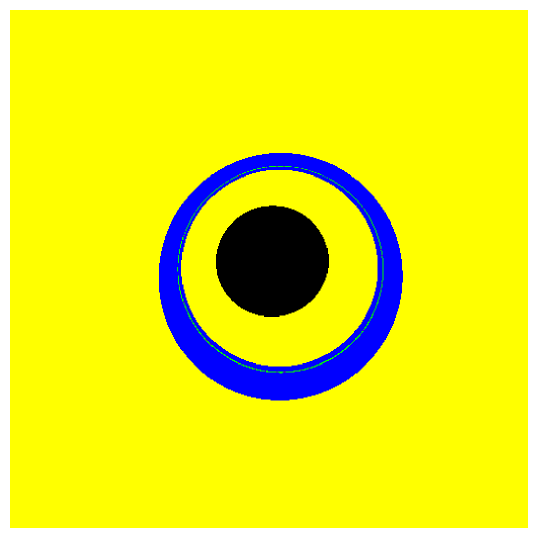}}
\subfigure[$a=0.99$, $\theta_o = 83^{\circ}$, $\xi=0.9$]{\includegraphics[scale=0.5]{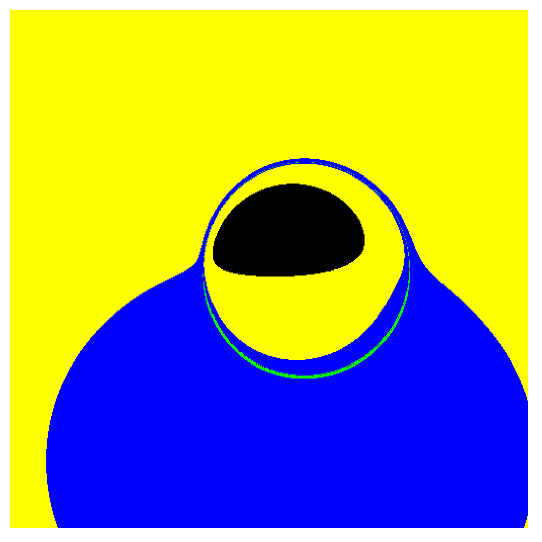}}
\caption{\label{figOB1}  Observed flux of the direct image, lensed image, and photon ring when the Konoplya-Zhidenko rotating non-Kerr black hole is surrounded by a prograde thin accretion disk. Here, the relevant parameters correspond to those in Figure \ref{figbpA1}. }
\end{figure}

To better distinguish the direct image, lensed image and photon ring region in the observed images at different observation inclinations, we present their observed fluxes in Figure \ref{figOB1}, where the parameter values correspond to those in Figure \ref{figbpA1}. In these images, yellow denotes direct images, corresponding to regions where light traverses the thin disk once; blue denotes lensed images, corresponding to regions where light passes through the thin disk twice; and green represents the photon ring, corresponding to regions where light passes through the thin disk three or more times. At low observation inclination angles ($\theta_o = 0^{\circ}$), the lensed image, direct image, and photon ring are all approximately circular in shape and are closely adjacent to one another. As the observational inclination angle increases, this axisymmetric circular feature is disrupted. When $\theta_o = 17^{\circ}$, the direct image and the lensed image exhibit gradual deformation, accompanied by a corresponding change in the morphology of the inner shadow region.  Crucially, at a high inclination of $\theta_o = 83^{\circ}$, the distortion is severe enough to spatially separate the direct and lensed images, allowing for their clear identification. Furthermore, the majority of the observed lensed emission flux is concentrated in the lower half of the observation plane, while only a minor fraction appears in the upper half. These results are consistent with the observed characteristics reflected in Figure \ref{figbpA1}.

\begin{figure}[!t]
\centering 
\subfigure[$a=0.3$,  $\theta_o = 0^{\circ}$]{\includegraphics[scale=0.6]{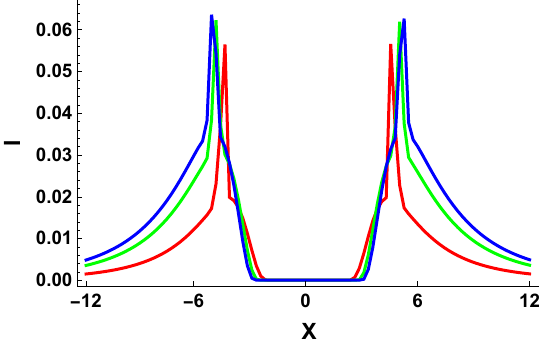}}
\subfigure[$a=1.15$,  $\theta_o = 0^{\circ}$]{\includegraphics[scale=0.6]{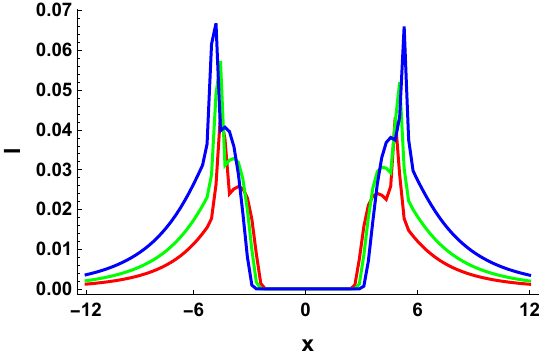}}
\subfigure[$\xi=0.5$,  $\theta_o = 0^{\circ}$]{\includegraphics[scale=0.6]{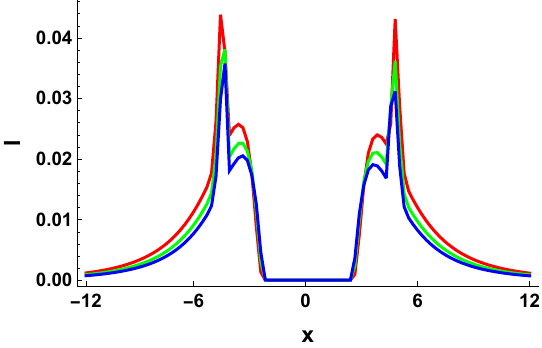}}
\caption{\label{figIB1} The intensity distribution under different parameters when the  observed inclination is taken as $\theta_o = 0^{\circ}$ and $M=1$. Plane (a): the red, green, and blue curves represent the cases of $\xi=-0.9, 0.1$ and $0.9$, respectively; Plane (b): the red, green, and blue curves correspond to $\xi=0.5, 1$ and $2$, respectively; Plane (c): the red, green, and blue curves correspond to $a = 1.05, 1.1$  and $1.15$, respectively.}
\end{figure}

On the other hand, changes in the parameter space and observation angles will affect the brightness of the image. The results show that the central brightness depression is a characteristic phenomenon observed in all cases where the accretion disk is not excessively thick. When the angle of observation is $\theta_o = 0^{\circ}$, the brightness of the image exhibits symmetry in both horizontal and vertical directions. However, as the angle of observation increases, the asymmetry of image brightness in the horizontal direction gradually intensifies. In particular, when the angle of observation is $\theta_o = 83^{\circ}$, there is a noticeable disparity in radiation flux intensity between the left and right sides of the image. A distinctly luminous, crescent-shaped region is visible on the left side of the image, resulting from Doppler effects. In the case of $a < M$, increasing both parameters $\xi$ and $a$ leads to an enhancement in the overall observed image intensity. When $a > M$, an increase in parameter $\xi$ results in higher observed intensity, whereas an increase in parameter $a$ leads to a reduction in the observed intensity. To more effectively demonstrate the impact of parameter variations on observation intensity, the intensity distribution along the $x$-axis is presented on the screen, see Figure \ref{figIB1}.  The results indicate that, in cases where $a > M$ and $a < M$, an increase in the deformation parameter enhances the observed image intensity. Conversely, an increase in the rotation parameter reduces the corresponding observed intensity.

For comparison with another important deformed non-Kerr black hole, we contrast the above Konoplya-Zhidenko thin disk images with those of the Johannsen--Psaltis (JP) black hole. The JP metric introduces a set of additional deviation parameters besides the mass and spin, and reduces to the Kerr metric when all deviation parameters vanish \cite{johannsen2011metric}. Bambi investigated the direct images of a geometrically thin and optically thick Novikov--Thorne accretion disk in the rotating JP spacetime using a ray-tracing method, showing that the JP deviation changes the morphology of the central dark region and the spatial distributions of the observed temperature and flux \cite{bambi2012code}. In comparison, our Konoplya-Zhidenko images show that increasing the deformation parameter $\xi$ enlarges the inner shadow. Recently, Olmo et al. compared thin disk images based on the leading-order terms of the spherically symmetric JP and Konoplya--Rezzolla--Zhidenko (KRZ) parametrizations under the same emission assumptions. Their results show that the JP and KRZ backgrounds have different effects on the central brightness depression and the shadow size, and also change the relative brightness of the photon ring \cite{olmo2025shadows}. These results indicate that the size of the central dark region and the photon ring brightness distribution may serve as important optical features for distinguishing the JP and Konoplya-Zhidenko non-Kerr spacetimes. It should be emphasized that the JP deviation parameters and the deformation parameter $\xi$ in the Konoplya-Zhidenko metric considered in this work are defined within different metric families, and that the accretion disk and emission prescriptions adopted in the related studies are not completely identical. Therefore, the above comparison should be regarded as qualitative. A strict quantitative comparison requires a unified model framework.

\begin{figure}[!t]
\centering 
\subfigure[$a=0.3$,  $\xi=-0.5$]{\includegraphics[scale=0.375]{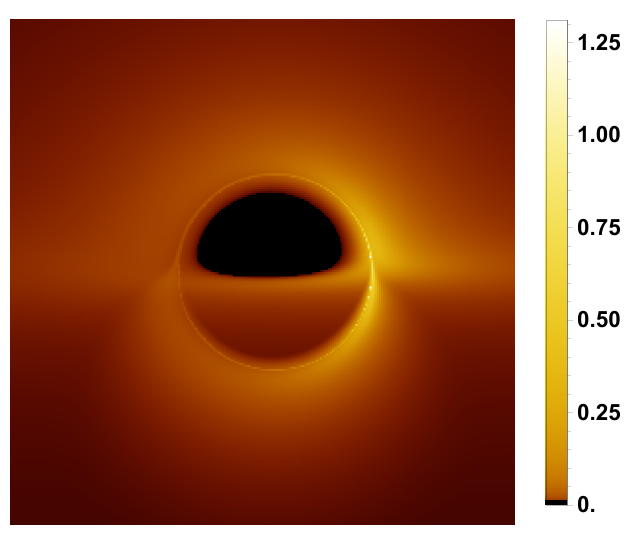}}
\subfigure[$a=0.3$,  $\xi=0.5$]{\includegraphics[scale=0.375]{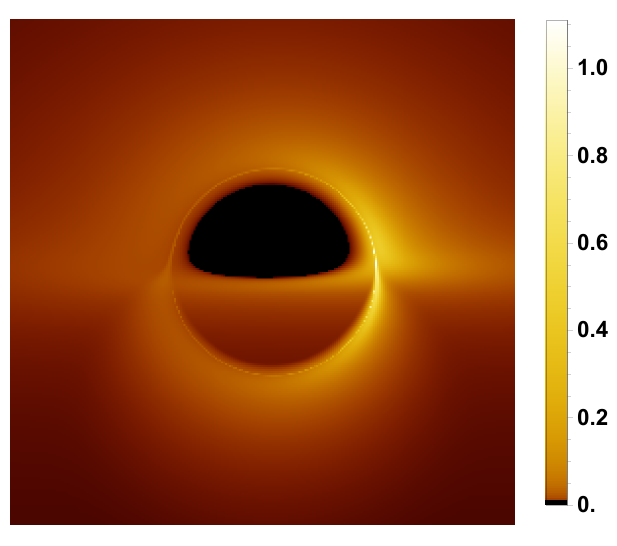}}
\subfigure[$a=0.6$,  $\xi=0.5$]{\includegraphics[scale=0.375]{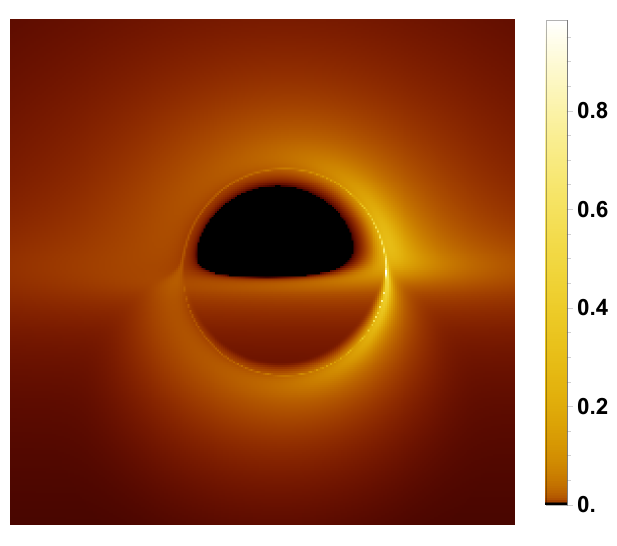}}
\subfigure[$a=1.15$,  $\xi=0.5$]{\includegraphics[scale=0.375]{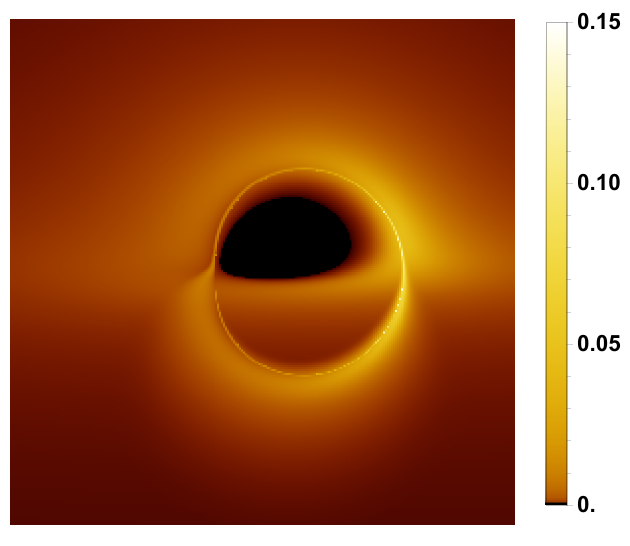}}
\caption{\label{figbpA3} The image of the Konoplya-Zhidenko rotating non-Kerr black hole surrounded by a retrograde thin accretion disk at 230 GHz, where the observed inclination is taken as $\theta_o = 83^{\circ}$.}
\end{figure}

We now proceed to further elucidate the optical observational characteristics of black holes surrounded by retrograde accretion disks. Figure  \ref{figbpA3} illustrates the impact of variations in spin parameters $a$ and deformation parameters $\xi$ on the image of a Konoplya-Zhidenko rotating non-Kerr black hole, which is surrounded by an accretion disk with retrograde flows. The first three columns correspond to the scenario where the spin parameter $a$ is less than the mass $M$ ($a<M$), while the fourth column pertains to the situation where the spin parameter $a$ exceeds the mass $M$ ($a>M$).  A consistent feature across all parameter configurations is the presence of the inner shadow region, underscoring its invariance with respect to the direction of accretion. As in the prograde case, an increase in the observational inclination angle facilitates a clear separation between the direct and lensed images. Moreover, variations in $a$ and $\xi$ exert analogous influences on the size and morphology of the inner shadow as observed in the prograde scenario, namely, an increase in $a$ reduces the shadow area, while an increase in $\xi$ enlarges it. However, a striking contrast emerges in the intensity distribution across the image plane. In retrograde accretion, the bright emission features, specifically the crescent-shaped structures, shift conspicuously to the right side of the image, as opposed to their left-side appearance in the prograde case. This reversal is a direct consequence of the opposite sense of movement of accreted matter relative to the black hole's spin.

\section{Distribution of the redshift factors}
The examination of how light from the accretion disk surrounding the black hole reaches the observer's plane necessitates careful consideration of variations in light intensity, which are influenced by a combination of divergence, absorption, Doppler effect, and gravitational redshift. The relative motion between the accretion disk and distant observers offers us a unique perspective into the intricate interactions between black holes and accretion disks, as well as the nature of their formation. The Doppler effect plays a significant role in explaining such phenomena, while the gravitational redshift, as a crucial indicator for revealing the impact of an extreme gravitational field, cannot be disregarded. Therefore, in the process of imaging a black hole, it is crucial to accurately assess the redshift factor associated with the behavior of emitted particles, as this step plays a critical role in precisely characterizing both the black hole and its accretion disk.
\begin{figure}[!h]
\centering 
\subfigure[$a=0.3$, $\theta_o = 0^{\circ}$, $\xi=-0.9$]{\includegraphics[scale=0.5]{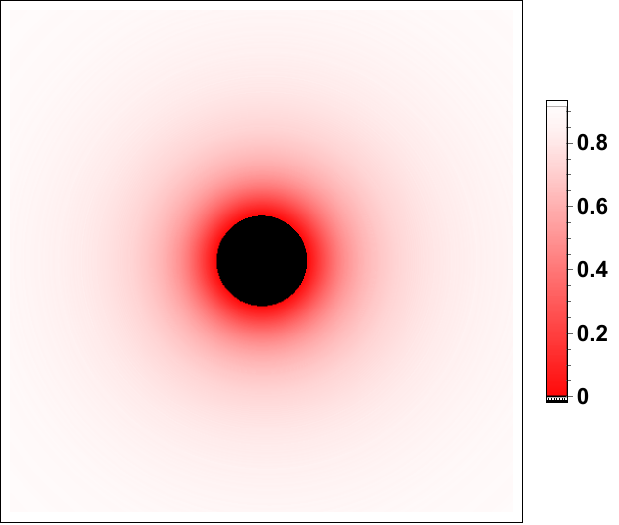}}
\subfigure[$a=0.3$, $\theta_o = 17^{\circ}$, $\xi=-0.9$]{\includegraphics[scale=0.5]{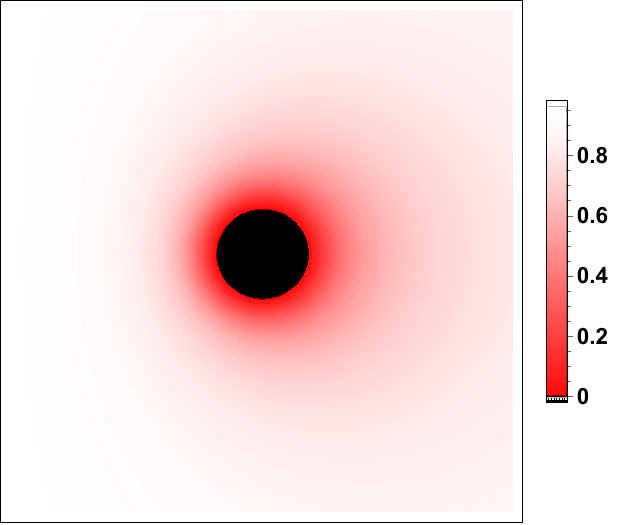}}
\subfigure[$a=0.3$, $\theta_o = 83^{\circ}$, $\xi=-0.9$]{\includegraphics[scale=0.5]{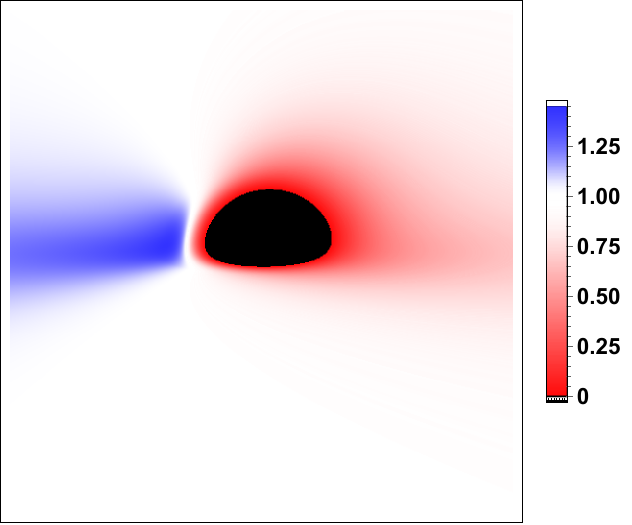}}
\subfigure[$a=0.3$, $\theta_o = 0^{\circ}$, $\xi=0.9$]{\includegraphics[scale=0.5]{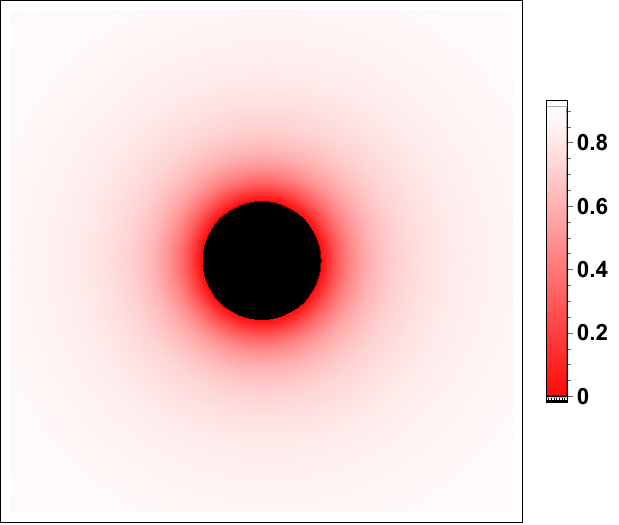}}
\subfigure[$a=0.3$, $\theta_o = 17^{\circ}$, $\xi=0.9$]{\includegraphics[scale=0.5]{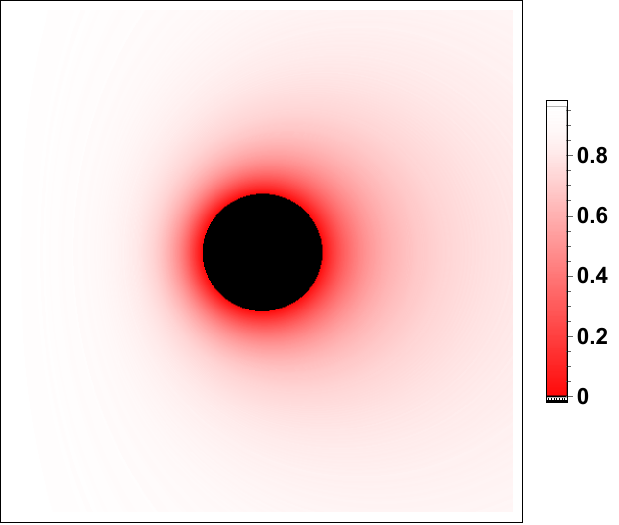}}
\subfigure[$a=0.3$, $\theta_o = 83^{\circ}$, $\xi=0.9$]{\includegraphics[scale=0.5]{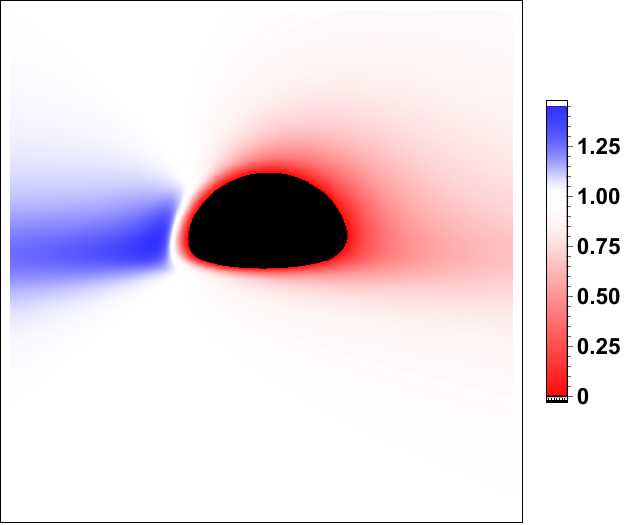}}
\subfigure[$a=0.99$, $\theta_o = 0^{\circ}$, $\xi=0.9$]{\includegraphics[scale=0.5]{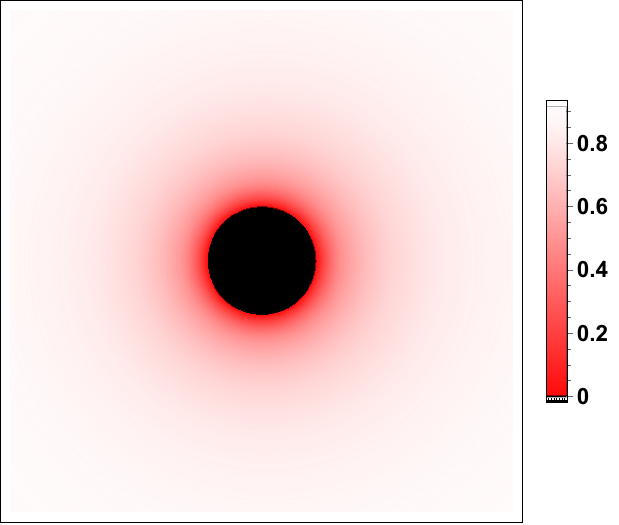}}
\subfigure[$a=0.99$, $\theta_o = 17^{\circ}$, $\xi=0.9$]{\includegraphics[scale=0.5]{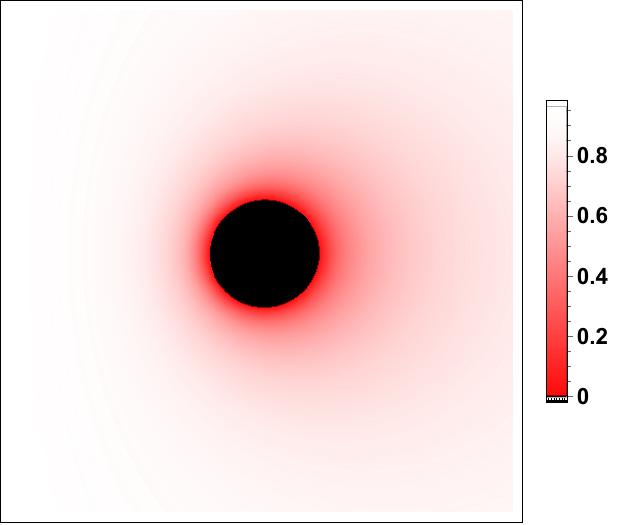}}
\subfigure[$a=0.99$, $\theta_o = 83^{\circ}$, $\xi=0.9$]{\includegraphics[scale=0.5]{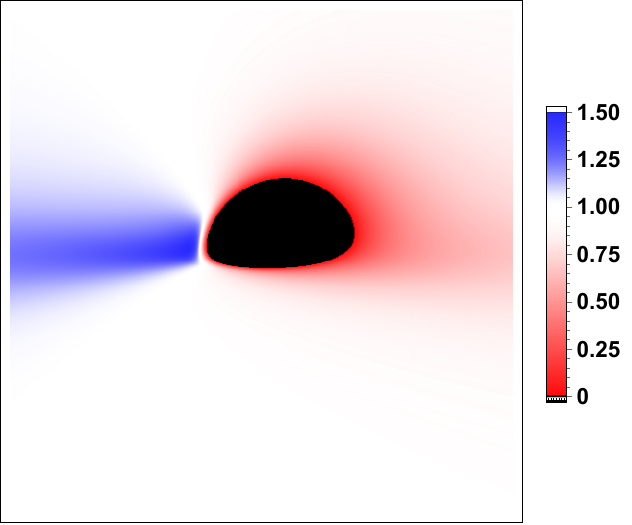}}
\caption{\label{figR1A1D} The redshift distribution of direct images under prograde disk, where the parameter values are taken as $a<M$. The red and blue hues denote the redshift and blueshift, respectively, while the shaded black region represents the event horizon of the black hole. }
\end{figure}

\begin{figure}[!t]
\centering 
\subfigure[$a=0.3$, $\theta_o = 0^{\circ}$, $\xi=-0.9$]{\includegraphics[scale=0.5]{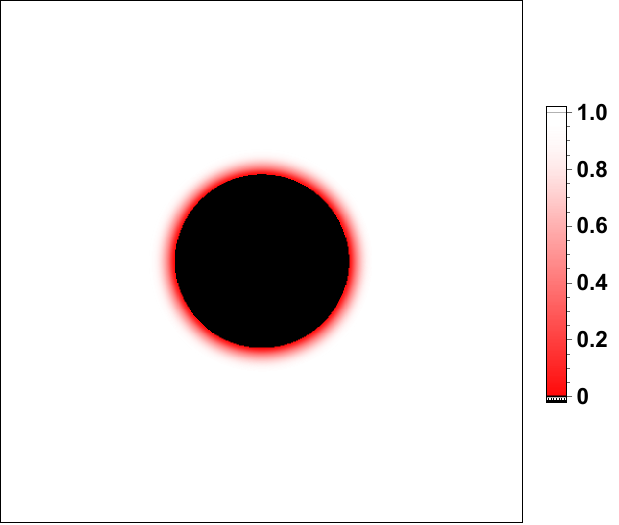}}
\subfigure[$a=0.3$, $\theta_o = 17^{\circ}$, $\xi=-0.9$]{\includegraphics[scale=0.5]{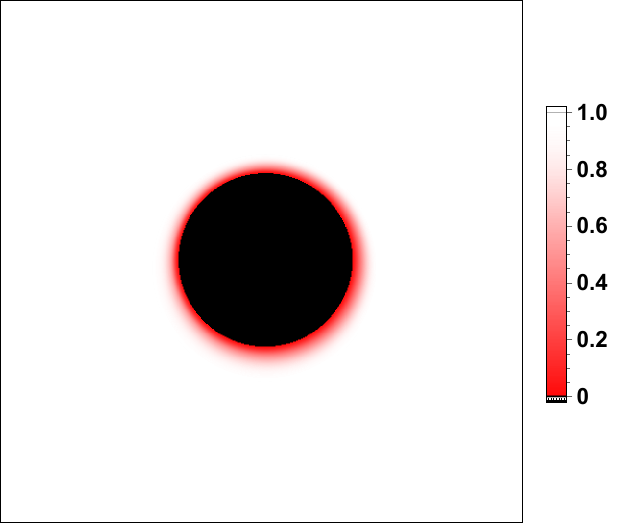}}
\subfigure[$a=0.3$, $\theta_o = 83^{\circ}$, $\xi=-0.9$]{\includegraphics[scale=0.5]{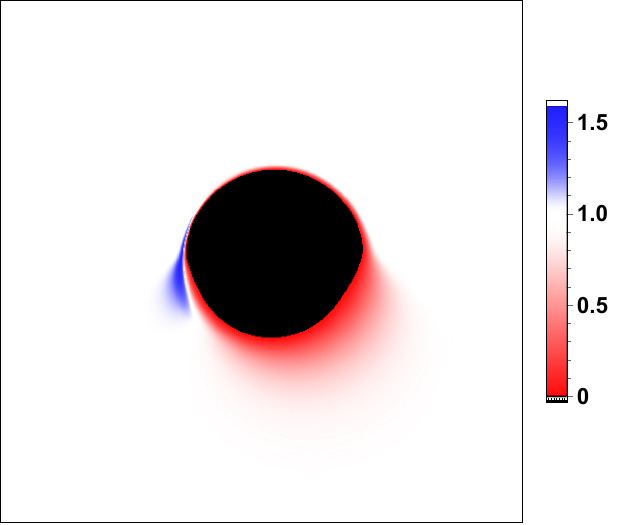}}
\subfigure[$a=0.3$, $\theta_o = 0^{\circ}$, $\xi=0.9$]{\includegraphics[scale=0.5]{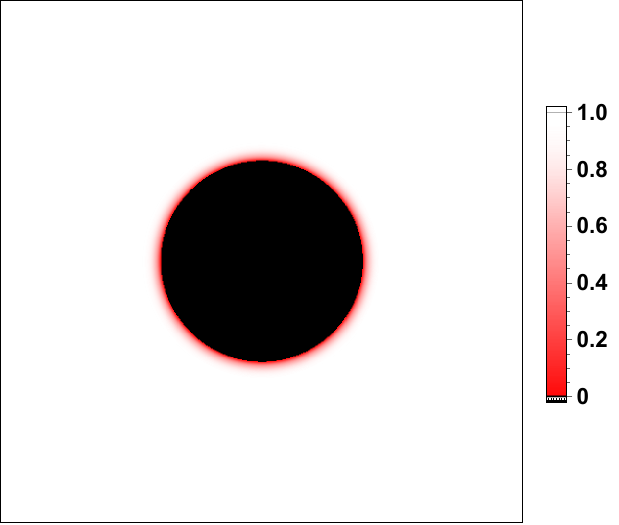}}
\subfigure[$a=0.3$, $\theta_o = 17^{\circ}$, $\xi=0.9$]{\includegraphics[scale=0.5]{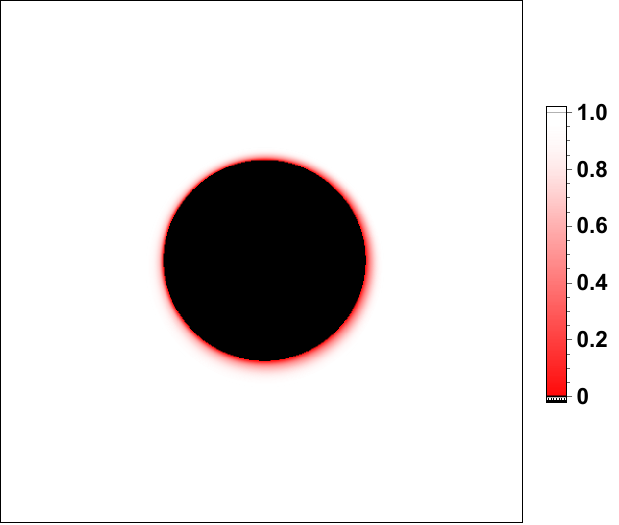}}
\subfigure[$a=0.3$, $\theta_o = 83^{\circ}$, $\xi=0.9$]{\includegraphics[scale=0.5]{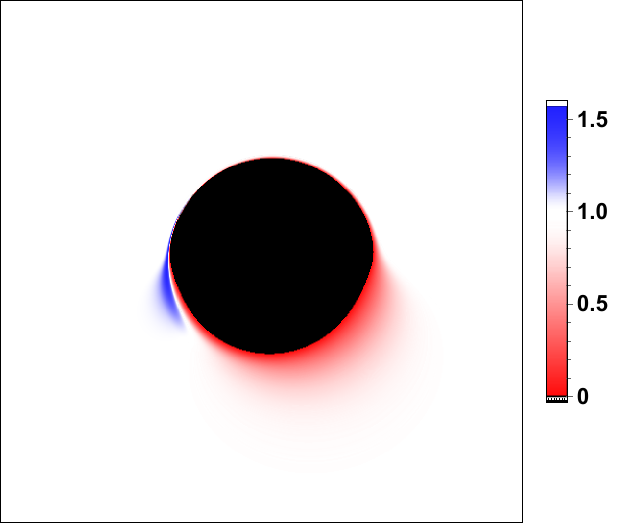}}
\subfigure[$a=0.99$, $\theta_o = 0^{\circ}$, $\xi=0.9$]{\includegraphics[scale=0.5]{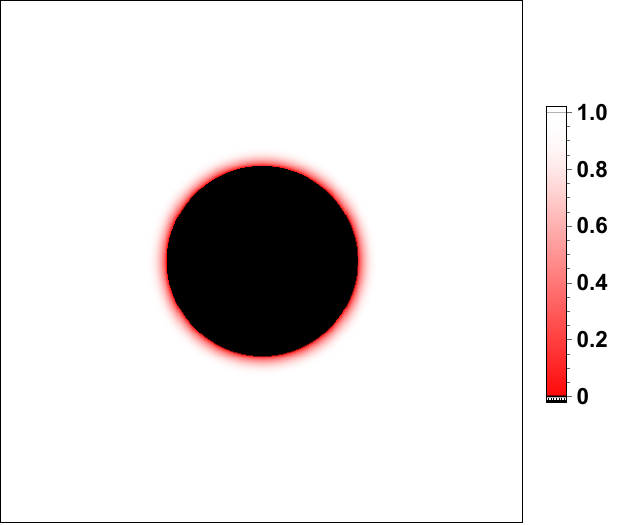}}
\subfigure[$a=0.99$, $\theta_o = 17^{\circ}$, $\xi=0.9$]{\includegraphics[scale=0.5]{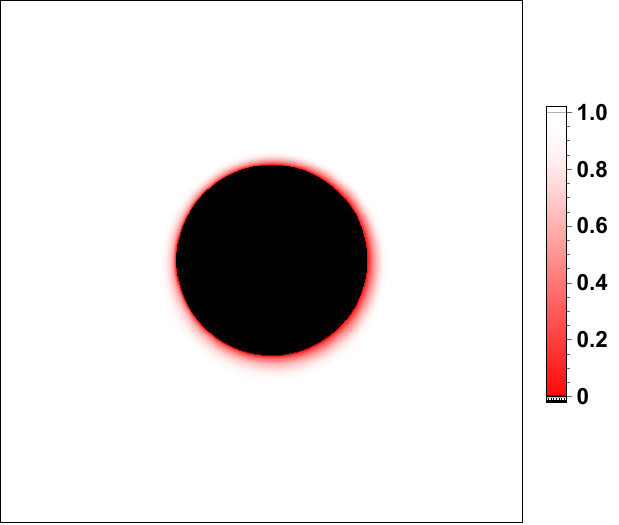}}
\subfigure[$a=0.99$, $\theta_o = 83^{\circ}$, $\xi=0.9$]{\includegraphics[scale=0.5]{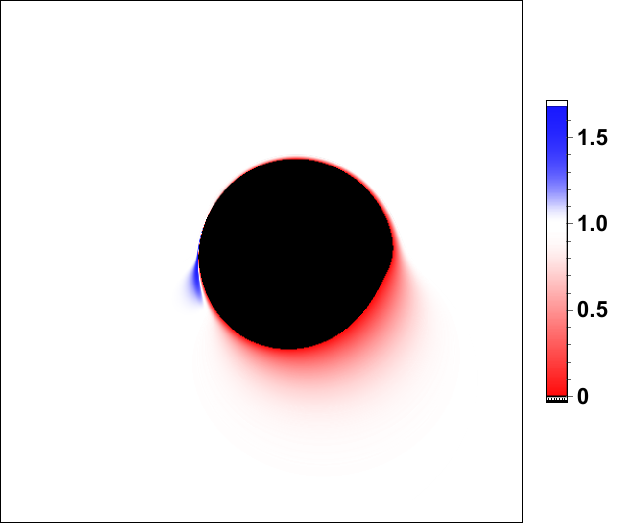}}
\caption{\label{figR2A1L} The redshift distribution of lensed images under prograde disk, where the parameter values are taken as $a<M$. The red and blue hues denote the redshift and blueshift, respectively, while the shaded black region represents the event horizon of the black hole. In addition, set the mass of the black hole to $M=1$ and fix the observer's position at $r_{obs}=500M$. }
\end{figure}

\begin{figure}[!t]
\centering 
\subfigure[$a=1.05$,  $\xi=0.5$]{\includegraphics[scale=0.5]{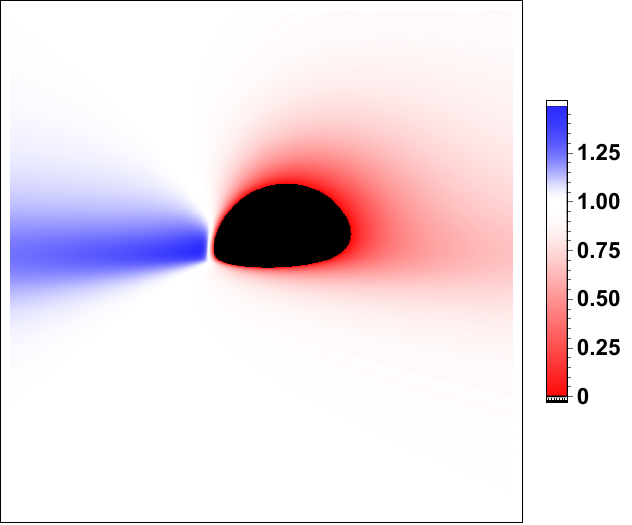}}
\subfigure[$a=1.05$,  $\xi=2$]{\includegraphics[scale=0.5]{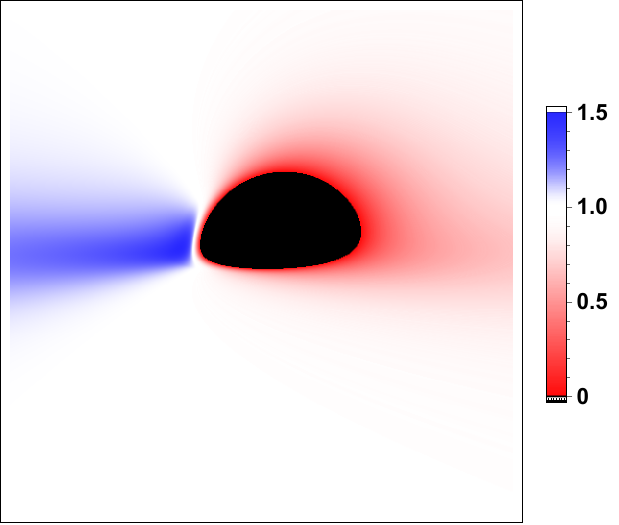}}
\subfigure[$a=1.15$,  $\xi=0.5$]{\includegraphics[scale=0.5]{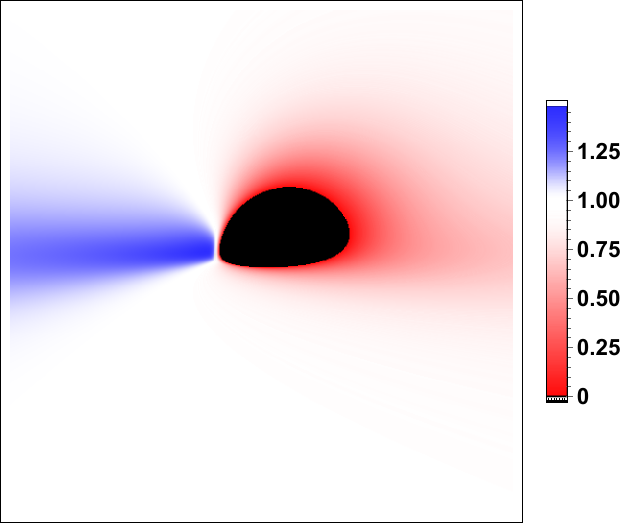}}
\subfigure[$a=1.05$,  $\xi=0.5$]{\includegraphics[scale=0.5]{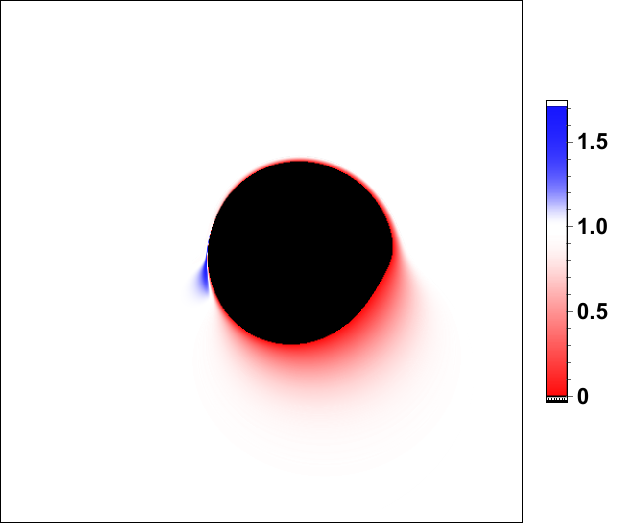}}
\subfigure[$a=1.05$,  $\xi=2$]{\includegraphics[scale=0.5]{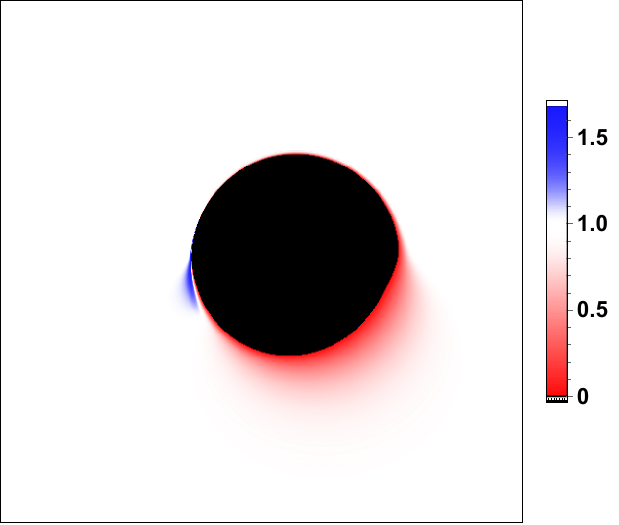}}
\subfigure[$a=1.15$,  $\xi=0.5$]{\includegraphics[scale=0.5]{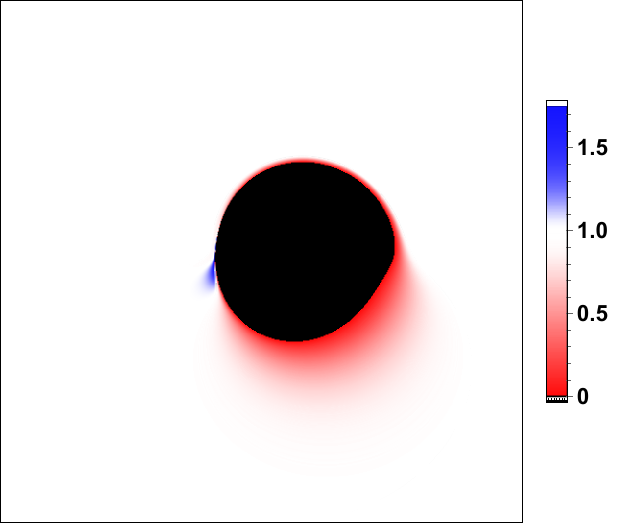}}
\caption{\label{figR1A2D} The redshift distribution of direct images under prograde disk, where the parameter values are taken as $a>M$. In addition, the observed inclination is $\theta_o = 83^{\circ}$  }
\end{figure}

In Figures \ref{figR1A1D} and \ref{figR2A1L}, we present the redshift factors of the direct and lensed images of the prograde accretion disk for the case where $a<M$,  while the other parameter values correspond to those in Figure \ref{figbpA1}.  A continuous linear color map is utilized to visually represent the redshift factor, with red indicating redshift and blue indicating blueshift. The central black region of each observing plane represents the inner shadow cast by the accretion disk, which is bounded by the projection of the event horizon of the black hole.  It can be observed that when the observed inclination value is small, such as $\theta_o = 0^{\circ}$ and $\theta_o = 17^{\circ}$, the observer's screen mainly displays a redshift feature. When the observation inclination is large, such as $\theta_o = 83^{\circ}$, an obvious blueshift also appears on the observer screen in addition to the redshift and is mainly distributed on the left side of the screen, while the redshift is mainly distributed on the right side. Furthermore,  an increment in  the deformation parameter $\xi$  results in a reduction of the redshift range while simultaneously expanding the blueshift range. Conversely, an augmentation of the rotational parameter $a$ leads to a substantial diminution in both the redshift and blueshift ranges. For the lensed image depicted in Figure \ref{figR2A1L}, a redshift is observed exclusively at low obliquity angles ($\theta_o = 0^{\circ}$ and $\theta_o = 17^{\circ}$), which aligns with the characteristics of the direct image. When the observed inclination is $\theta_o = 83^{\circ}$, a minor petal-shaped blueshift appears on the lower right side of the screen, while the redshift remains in proximity to the ISCO. Furthermore, an increase in the deformation parameter $\xi$ enlarges the size of the blue-shifted shape, while an increase in the parameter $a$ has the opposite effect.
For the case of $a > M$, the redshift factors of the direct (first row) and lensed (second row) images of  the prograde accretion disk  are presented in Figures \ref{figR1A2D} when the observed inclination is $\theta_o = 83^{\circ}$. It can be observed that as the deformation parameter $\xi$ increases or the rotation parameter $a$ decreases, the blue-shifted region gradually expands, especially near the ISCO, where the blueshift phenomenon is more pronounced.

\begin{figure}[!t]
\centering 
\subfigure[$a=0.3$, $\xi=-0.5$]{\includegraphics[scale=0.375]{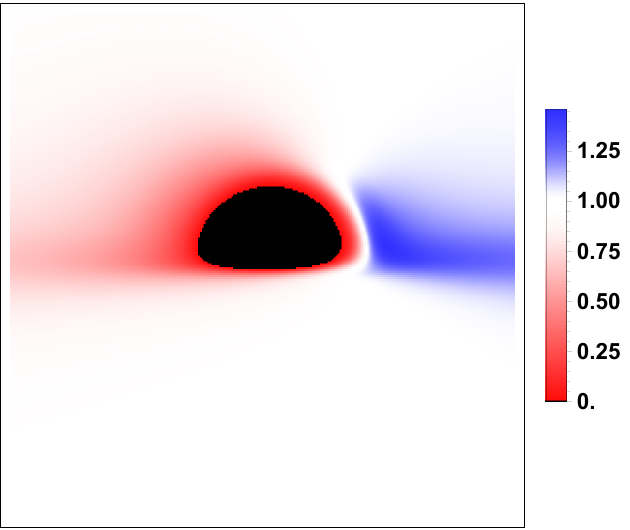}}
\subfigure[$a=0.3$,  $\xi=0.5$]{\includegraphics[scale=0.375]{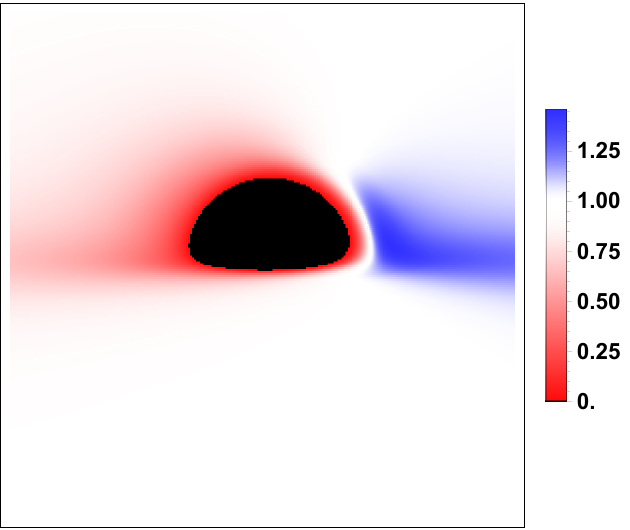}}
\subfigure[$a=0.6$,  $\xi=0.5$]{\includegraphics[scale=0.375]{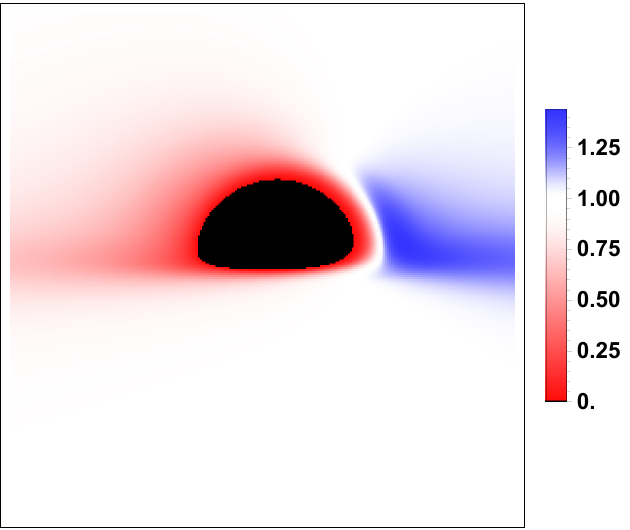}}
\subfigure[$a=1.15$,  $\xi=0.5$]{\includegraphics[scale=0.375]{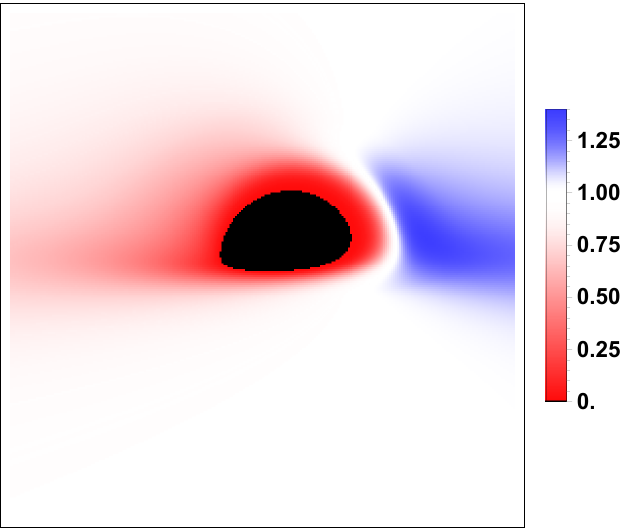}}
\subfigure[$a=0.3$,  $\xi=-0.5$]{\includegraphics[scale=0.375]{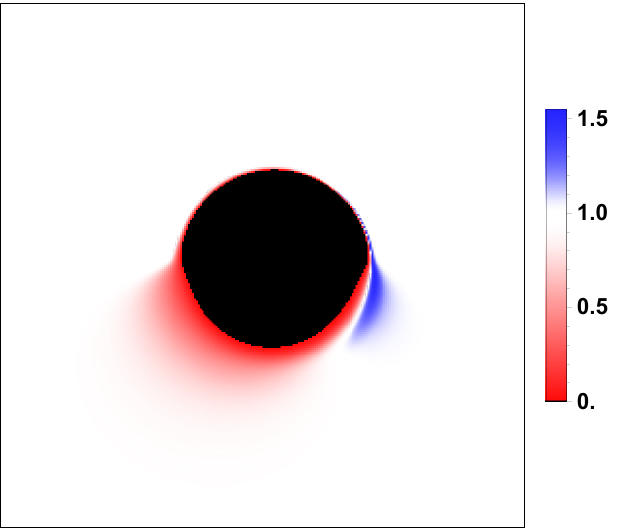}}
\subfigure[$a=0.3$,  $\xi=0.5$]{\includegraphics[scale=0.375]{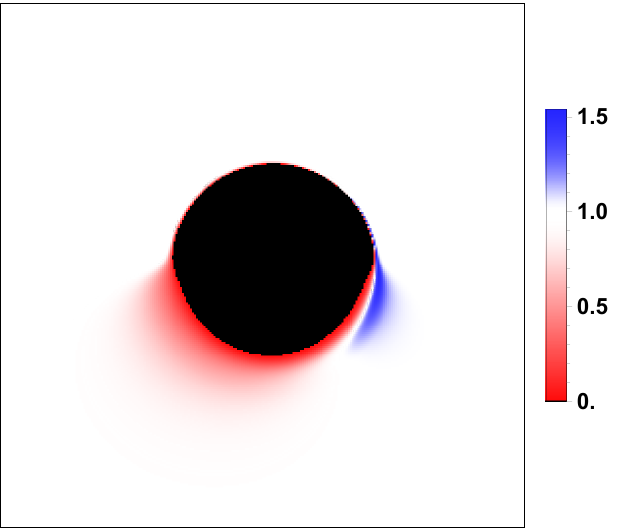}}
\subfigure[$a=0.6$,  $\xi=0.5$]{\includegraphics[scale=0.375]{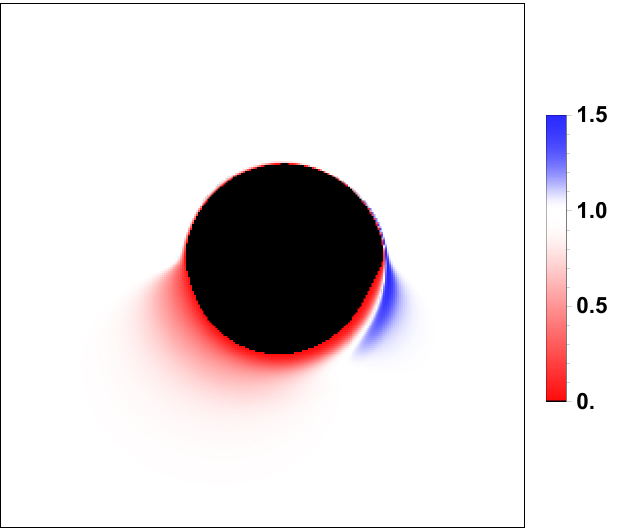}}
\subfigure[$a=1.15$,  $\xi=0.5$]{\includegraphics[scale=0.375]{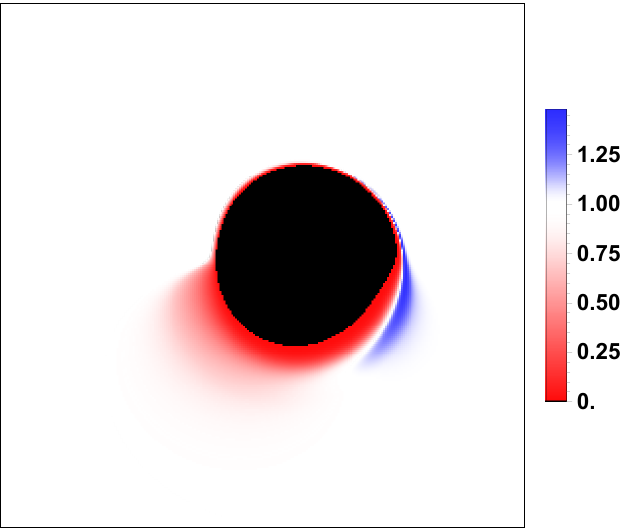}}
\caption{\label{figNARO83} The image of a Konoplya-Zhidenko rotational non-Kerr black hole  surrounded by a retrograde  thin accretion disk, where the observed inclination is $\theta_o = 83^{\circ}$  }
\end{figure}

Figure \ref{figNARO83} depicts the redshift distribution for retrograde accretion at an inclination of $\theta_o = 83^{\circ}$, showcasing the direct image (top row) and the lensed image (bottom row). A key observation is that the redshift and blueshift regions are reversed compared to the prograde case (Figures \ref{figR1A1D} and \ref{figR2A1L}). This inversion is a direct consequence of the reversed Doppler effect, that is, the accretion flow now moves counter to the black hole's spin, causing the previously blueshifted (bright) approaching side to appear on the opposite side of the image.
For the direct image, the blueshifted region (now on the right) expands as the deformation parameter $\xi$ and the spin parameter $a$ increase. The lensed image, however, exhibits only an arc-like blueshifted structure confined to the lower right corner. The lensed image is concentrated on the lower right side of the screen, which is consistent with the observed intensity distribution shown in Figure \ref{figbpA3}, that is, the main observed intensity is mainly concentrated on the lower half of the screen.

\section{Conclusions and discussions}
\label{sec:conclusion}
The formation of a black hole shadow is fundamentally due to the deflection of light in an intense gravitational field, and its image can provide crucial insights into the geometry of spacetime. The Konoplya-Zhidenko rotational non-Kerr spacetime has some non-negligible deviations from the Kerr spacetime, and the quasi-periodic oscillations and the constraints of iron lines also support that a real astrophysical black hole can be described by the Konoplya-Zhidenko rotational non-Kerr metric\cite{Bambi:2012pa,Bambi:2013fea,Kong:2014wha,Bambi:2011ek,Johannsen:2012ng,Jiang:2014loa,Bambi:2013sha}. In this paper, we investigate the shadow and optical observational properties of a Konoplya-Zhidenko rotating non-Kerr metric black hole to explore the observed differences between different parameters.  As an initial step, we investigate the dynamics of photons in the vicinity of the Konoplya-Zhidenko rotational non-Kerr black hole and constrain the relevant parameters of this spacetime based on astronomical observations of M87* and Sgr A*. The results indicate that an increase in $\xi$ effectively enhances the gravitational strength in the near-horizon region, thereby expanding the photon capture cross-section and leading to a larger shadow size (with $R_d$ increasing) compared to the corresponding Kerr black hole with the same spin parameter $a$.

To acquire the observational image of the black hole from the perspective of a ZAMO, we propose positioning an optically and geometrically thin accretion disk on the equatorial plane of the black hole, serving as the sole background light source. We extend the classical accretion disk model to a scenario in which the innermost region of the accretion disk extends to the event horizon of the black hole. Furthermore, the accretion materials consist of electrically neutral plasma, and their motion within the accretion disk is categorized into two types based on the ISCO. Specifically, the ISCO delineates the boundary beyond which the emission spectrum from accretion disks will exhibit stable circular orbits. Within this boundary, however, material is subject to critical plunging orbits. Using fisheye camera ray-tracing techniques, we conducted a detailed analysis of the image of a thin accretion disk surrounding a Konoplya-Zhidenko rotating non-Kerr black hole, where the motion of the accretion disk in two ways, i.e., prograde and retrograde.

Firstly, we investigate the image of a Konoplya-Zhidenko rotating non-Kerr metric black hole encompassed by a prograde thin accretion disk, see Figure \ref{figbpA1} and Figure \ref{figbpA2}. In all these images, one can consistently observe a prominent dark region surrounded by a narrow photon ring. The dark region corresponds to the accretion disk profile at $r = r_h$, commonly referred to as the inner shadow, while the photon ring aligns closely with the critical curve of the black hole. Regardless of whether  $a>M$ or $a<M$, an increase in the deformation parameter $\xi$ will expand the area of the inner shadow region. At low observation angles ($\theta_o = 17^{\circ}$), although the shadow region in $\theta_o = 17^{\circ}$ has a slight deformation and moves to the upper left of the image screen, it does not show obvious photon ring concentric circles like $\theta_o = 0^{\circ}$, but the difference in brightness distribution near the image is not obvious, so that the direct image and the lens image cannot be directly distinguished. When the observed inclination is $\theta_o = 83^{\circ}$, the deformation of the inner shadow region becomes clearly identifiable, exhibiting a shape reminiscent of a hat. Then, a clear distinction can be made between the direct image and the lensed image. It should be noted that a bright crescent-shaped or eyebrow-like region is visible on the left side of the screen, as illustrated in Figure \ref{figbpA1} and Figure \ref{figbpA2}, which results from the Doppler effect. To accurately analyze the real situation of the accreted matter around the black hole, we also examined the redshift distribution of the direct image and the lensed image. In the case of direct images, the results indicate that at a smaller observation angle($\theta_o = 17^{\circ}$), the observer can only detect the redshift factor on the screen. As $\theta_o$ increases to $\theta_o=83^{\circ}$, a blueshift is also observed on the left side of the screen. In the case of the  lensed image, the observed inclination changes are analogous to those observed in the direct image. However, a notable distinction is the petal-like blueshift that appears on the left side of the screen when $\theta_o = 83^{\circ}$, which is considerably smaller in magnitude compared to the direct image results. Furthermore, the alteration of the parameter space will result in modifications to the ranges of redshift and blueshift.

Subsequently, we examine the image of a Konoplya-Zhidenko rotating non-Kerr  black hole, wherein the accretion disk exhibits retrograde motion, see Figure\ref{figbpA3}. It is evident that irrespective of the orientation of the accretion disk, the black hole's intrinsic shadow remains invariant under identical parameters. Likewise, when the angle of observation is sufficiently large, it becomes feasible to differentiate between the direct image and the lensed image. Undoubtedly, there are distinct differences in the imagery of prograde and retrograde accretion disks. In the case of a retrograde accretion flow, the observed intensity distribution on the screen is predominantly concentrated on the right-hand side, which is in direct contrast to the prograde scenario. This phenomenon occurs because in the case of reverse flow, the movement direction of the accreting matter is opposite to the spin direction of the black hole, and the Doppler effect changes accordingly. This result is also evident in the redshift distribution. In the case of retrograde accretion, the blueshifted region appears on the right side of the image, which is precisely opposite to that observed in prograde accretion.

By analyzing the shadow and optical characteristics of black holes, it is indeed possible to extract certain characteristic information regarding the deformation parameter $\xi$ in the Konoplya-Zhidenko rotational non-Kerr spacetime. The optical observational characteristics of black holes are closely associated with the relevant parameter space, the observational inclination angle, and the motion behavior of the accreted matter. In the present work, we have restricted our analysis to black hole configurations admitting an event horizon. The optical appearance of naked singularity spacetimes, including cases with and without photon spheres, may differ qualitatively and deserves a separate systematic investigation in future work. We anticipate that this work will contribute to the theoretical research of high-resolution accretion disk imaging in the future, and offer valuable inspiration for validating general relativity and other modified gravitational theories.

\section*{Data Availability Statement}
\noindent The data that support the findings of this study are available from the corresponding author upon reasonable request.

\vspace{10pt}

\noindent {\bf Acknowledgments}

\noindent
This work is supported by the National Natural Science Foundation of China (Grants Nos. 12505059, 12375043, 12575069), and the China Postdoctoral Science Foundation (Grants No.2025MD784184), and Chongqing Normal University Fund Project (Grants No. 26XLB001).



\end{document}